# Title: Handed Ultrafast Dynamics in Chiral Cyanine Nanotubes

**Authors:** Indranil Ghosh[1,†], Keerthi Vaasan Mani[1,†], Caitlin C. Bellora[§], Gregory S. Engel[1,*]

**Affiliations:**

[1]Department of Chemistry, Institute for Biophysical Dynamics, James Franck Institute, and Pritzker School of Molecular Engineering at The University of Chicago; Chicago, IL 60637, USA

[†]Contributed equally

*Corresponding author: gsengel@uchicago.edu

**Abstract:**

We detect helicity-dependent excited state relaxation in a chiral exciton system. Using L-histidine as a chiral inducer, we prepared unbundled, double-walled J-aggregrate nanotubes of an amphiphilic cyanine dye with a net handedness, which gives a large circular dichroism dissymmetry factor ($g_{abs} = 3.6 \times 10^{-2}$). With two-dimensional electronic spectroscopy and circularly polarized pump pulses, we find that the relaxation rates of longitudinal excitons depend on pump helicity (4.5 ps$^{-1}$ for $\langle RRXX \rangle$ vs 6 ps$^{-1}$ for $\langle LLXX \rangle$, a difference of about 30%) and is the same in the inner and outer tubes. Transverse excitons show no significant dependence. Chirality and pump helicity therefore bias exciton relaxation on the femtosecond timescale for excitons delocalized along the helical structure.

**Main Text:**

Chiral molecules exhibit differential electronic absorption of left- versus right-handed circularly polarized light, a phenomenon known as electronic circular dichroism (ECD) (*1*). ECD reports on the chirality of the electronic ground state, and is widely used as a sensitive structural probe for biology (*2*). Circularly polarized luminescence (CPL) shows that excited states can also be chiral (*3*), but cannot reveal the chiral dynamics that precede the asymmetric emission of photons. Extending chiral chemical control to electronic excited states could enable stereoselective photochemistry inaccessible from the ground state (*4*) and materials that program the angular momentum of light (*5*). Here, we ask whether chirality also biases the excited state dynamics after light is absorbed, and whether light can direct this bias. Such control would open routes to asymmetric photocatalysis (*6*) and chiroptical information processing (*7*).

Two-dimensional electronic spectroscopy (2DES), which resolves both excitation and emission energies with femtosecond time resolution, has tracked exciton dynamics in photosynthetic complexes (*8*, *9*), small organic molecules (*10*, *11*), molecular materials (*12*) and solid-state systems (*13*, *14*). These studies typically used polarization sequences that are insensitive to handedness in isotropic samples. 2D circular dichroism has revealed chirality-dependent excitonic dynamics in the bacterial LH2 antenna (*15*), and chirality-induced spin selectivity (CISS) has been observed in photoinduced electron transfer in donor-acceptor molecules (*7*). However, the protein's complexity obscures interpretation, and molecular dyads lack the delocalized excitons of natural antennae.

Self-assembled molecular J-aggregates, resulting from the end-to-end arrangement of molecular transition dipoles (*16*), are an ideal platform for resolving chiral electronic dynamics in the ultrafast timescale. They are archetypal systems for excitonic chirality, as helical coupling of strongly coupled transition dipoles yields delocalized excitons with intense ECD (*17*, *18*). The lowest energy state in the excited state manifold of a J-aggregate is bright, and its electric transition dipole moment increases by $\sqrt{N}$ when the exciton delocalizes across $N$ monomers (*16*). Together, these properties yield narrow spectral lineshapes that resolve multiple spectroscopic features close in energy (*16*). We prepare non-racemic, unbundled double-walled C8S3 nanotubes using the chiral inducer L-histidine, which previous attempts could not achieve, and resolve their exciton dynamics with polarization-controlled 2DES. Pumping with opposite circular polarizations produces different relaxation rates for the longitudinal excitons demonstrating that the interplay of structural chirality and the polarization of light biases relaxation for excitons delocalized along the helical structure. These results reveal chiral selectivity of exciton dynamics.

**Inducing chirality in a double-walled nanotube ensemble**

Investigating the chiral excited states of supramolecular assemblies requires samples enriched in one handedness. Double-walled J-aggregate nanotubes of the amphiphilic cyanine dye 3,3'-bis (3-sulfopropyl)-5,5',6,6'-tetrachloro-1,1'-dioctylbenzimidacarbocyanine (C8S3) (*19*–*21*) are synthetic analogs of natural light harvesting systems. Their self-assembly has been characterized extensively, making them a well-established model system (*20*, *22*, *23*). The chemical structure of the C8S3 monomer is shown in **Fig. 1C**. The aggregates, which consist of two concentric cylindrical nanotubes, show superradiance and high coherence lengths of tens of monomers (**Fig. 1D**) (*24*). High-resolution cryo-transmission electron microscopy (TEM) showed that individual nanotubes are helical, yet no consistent ECD signal was observed (*20*), indicating a racemic population in which left- and right-handed nanotubes give cancelling signals. Previous efforts to obtain a non-racemic mixture of C8S3 and C8O3 nanotubes caused extensive nanotube bundling, which broadens the absorption spectrum and obscures excitonic features (*25*–*27*). Using the chiral inducer L-histidine, we prepare unbundled, double-walled C8S3 nanotubes with a net handedness (**Fig. 1**), making their chiral excited states accessible to chiroptical spectroscopy.

We chose L-histidine as the chiral inducer for two reasons: First, C8S3 is negatively charged, which favors an attractive ionic interaction the positively charged side chain of L-histidine. Second, both C8S3 and L-histidine contain an imidazole moiety, which can interact through π-orbital overlap, biasing the bricklayer stacking (*28*) of the C8S3 monomers that constitute the helical nanotube walls. An L-histidine/C8S3 sugar solution was rapidly drop-cast onto a quartz substrate, forming a sugar matrix that kinetically traps the aggregates in double-walled form and prevents bundling, (*19*, *29*). A complete synthetic protocol is provided in **section S1**. Hereafter, we call the non-racemic aggregate "chiral" for brevity.

The absorption spectra show bands at 599 nm and 587 nm (**Fig. 1B**), corresponding to the longitudinal excitons of the inner and outer tubes, respectively, and cryo-TEM images (**Fig. 1E**, **fig. S3**) confirm a double-walled nanotubular morphology (*19*, *20*, *23*). Previous work showed that bundling eliminates the outer-tube signature at 587 nm and produces a broad shoulder centered near 575 nm (*23*), so the retained 587 nm band and the absence of a shoulder indicate that the nanotubes were unbundled. The ECD spectrum shows corresponding bands at these wavelengths, indicating that the prepared sample is enriched in one handedness. We observe a chiral dissymmetry factor ($g_{abs}$) of $3.6 \times 10^{-2}$ at both 587 nm and 599 nm (**section S2**). This $g_{abs}$ is considered large for self-assembled supramolecular systems (*30*) and especially notable for long visible wavelength transitions, where $g_{abs}$ is typically diminished because it scales inversely with wavelength (*3*). A detailed analysis and discussion of the electronic structure of C8S3 nanotubular aggregates in **section s3**.

## Longitudinal excitons show chiral dynamics

Self-assembled C8S3 nanotubes combine a helical supramolecular structure with well-characterized delocalized excitons, providing a simple, well-defined platform for chiral exciton dynamics. Our goal is to show that pump helicity selectively biases these dynamics. 2DES resolves both excitation and emission energies, allowing us to track multiple electronic excited state populations with femtosecond time resolution. Previous 2DES studies of racemic C8S3 double-walled nanotubular aggregates used the all-parallel $\langle XXXX \rangle$ pulse sequence to resolve energy transfer and coupling between the inner and outer tubes (*23*), exciton-exciton annihilation (*31*), and lineshape broadening (*32*). The standard polarization sequence $\langle XXXX \rangle$ responds identically to left and right-handed structures (*33*), so it cannot reveal chirality-dependent dynamics for an isotropic sample like our aggregates. Transient CD spectroscopy, theoretically pioneered by Cho (*34*) and Mukamel (*35*), showed that odd sequences of $X$-and $Y$-polarized pulses in four-wave mixing (e.g., $\langle YXXX \rangle$, $\langle XXYX \rangle$, etc.) eliminate purely electric dipolar contributions, leaving magnetic dipolar and electric quadrupolar contributions to dominate the signal. Building on this, we develop a 2D electronic spectrometer that uses circularly-polarized pump pulses, linearly polarized probe and local oscillator pulses to control, not only probe, exciton relaxation dynamics in chiral nanotubes (**Fig. 2A**) (*36*). We chose 2DES over transient absorption because the excitation energy axis separates the overlapping spectral features (*31*) of the inner and outer tube excitons (**Fig. 2**), letting us resolve each tube's contribution to the chiral dynamics. We use $\langle LLXX \rangle$ and $\langle RRXX \rangle$ pulse sequences, where '$L$' and '$R$' denote left and right circular polarizations. Their difference contains only odd X/Y sequences, so any difference between them reflects the chiral response (**section s4**).

Representative 2DES spectra for the two circular-polarization sequences, $\langle LLXX\rangle$ and $\langle RRXX\rangle$, are nearly identical (**Fig. 2B**, **fig. S7-S9**) and match with previously reported 2D spectra of the racemic nanotubes (*21*, *31*), showing that L-histidine and circular polarization do not alter the spectral fingerprint. Despite this, single exponential fits to the excited state absorption (ESA) traces reveal significantly different relaxation rates between the two sequences, for both the inner and the outer tube longitudinal excitons (**Fig. 3**, fitting and statistical analysis in **section S5**) (*37*, *38*). The ESA signal decays with a 4.5 $ps^{-1}$ rate constant for $\langle RRXX\rangle$ and a 6 $ps^{-1}$ rate constant for $\langle LLXX\rangle$ for both the inner and outer tubes (F-test, $p_{inner} = 7.0 \times 10^{-6}$ and $p_{outer} = 9.0 \times 10^{-4}$; **Fig. 3**). Measurements at 0.8 nJ per pulse ensured negligible contribution from fifth-order signals (*31*), and an achiral laser dye Nile Blue under the same optical alignment show no difference between left- and right-handed excitations (**section S6**), indicating a molecular origin.

Pumping with opposite helicities therefore produces distinct exciton relaxation dynamics within a picosecond (**Fig. 3, fig. S13**). This overall 4-6 $ps^{-1}$ longitudinal exciton relaxation component was not resolved in racemic aggregates (*31*, *39*) . We attribute this fast relaxation to coupling to intra- and inter-molecular vibrations, consistent with a Stokes shift of 55 $cm^{-1}$ in the fluorescence spectrum (**fig. S2**), larger than that reported for racemic aggregates (*19*).

Longitudinal excitons, polarized along the tube axis, dominate absorption and are spectrally resolved for the inner and outer tubes, providing a well-characterized baseline (*21*, *31*). Transverse excitons, polarized around the tube circumference have much less oscillator strength than their longitudinal counterparts, the latter being far more delocalized (*40*, *41*). The inner tube transverse exciton at about 565 nm (*20*) is the only well-resolved feature in the absorption spectrum (**fig. S4**). While this excitation falls within our laser bandwidth (**Fig. 1B**), it is too weak to appear on the diagonal of the 2D spectrum, where excitation and emission energies match, and which shows no features bluer than 570 nm (**Fig. 2**). We do observe a positive below-diagonal cross-peak with excitation at 565 nm and emission at the inner tube longitudinal exciton energy at about 598 nm. Since this cross-peak shares the oscillator strength of the longitudinal exciton, it is above noise, and its evolution with $T$ is the only feature we can use to track transverse inner tube exciton dynamics.

Excitation of the inner tube transverse exciton is followed by an ultrafast energy transfer to the longitudinal exciton, seen as a rise in cross-peak amplitude with $T$ (**Fig. 4**). The sub-100 fs transfer time does not depend on pump polarization. The rates of energy transfer ($k_r$) between $\langle LLXX\rangle$, $\langle XXXX\rangle$ and $\langle RRXX\rangle$ pulse sequences show no significant difference (**Fig. 4D**, P = 0.25). The cross-peak amplitude does differ between the circular-pump and linear-pump pulse sequences, but this reflects the polarization dependence of signal strength for nonparallel transition dipoles (*42*), not helicity-dependent dynamics, because the rate is unchanged (**section S5**). In contrast to the longitudinal excitons, therefore, the transverse excitons show no detectable helicity dependence. This is consistent with their weak ECD signal at ~565 nm (**Fig. 1, fig. S5, table S2**), indicating weak rotational strength for the transverse transition dipole moment. Longitudinal transition dipoles are significantly more delocalized along the helical arrangement of the nanotube, a chiral character that may account for their chiral dynamics.

**Discussion**

We have shown that the relaxation of longitudinal excitons in non-racemic C8S3 nanotubes depends on pump helicity, with rates of 4.5 and 6 $ps^{-1}$ for right- and left-circularly polarized excitation (a difference of about 30%) that are the same within error for the inner and outer tubes. In contrast, transverse excitons show no significant helicity dependence, which is consistent with the static chiroptical response. Longitudinal excitons have intense ECD, whereas transverse excitons have weak ECD at ~565 nm, indicating low rotational strength. We propose that the helicity-dependent relaxation of the longitudinal excitons reflects their much greater delocalization along the helical arrangement, as an exciton spanning more of the helix couples more strongly to its handedness (*40*, *41*).

Identifying the mechanism behind the helicity-dependent relaxation requires the energies and wavefunctions of the chiral exciton states. ECD, unlike absorption and linear dichroism, is sensitive to the handedness of the molecular packing, so our ECD spectrum provides an independent constraint on the Hamiltonian that describes them. A Hamiltonian consistent with the absorption and linear dichroism spectra of the double-walled nanotubes and with their cryo-EM structure, has yet to be established (*20*, *28*, *40*, *44*) We outline two possible mechanisms. In Redfield theory, which describes relaxation through weak coupling to a vibrational bath, the exciton state preferentially excited by left-circularly polarized light may lie at an energy where the bath spectral density is larger, accelerating its relaxation relative to its right-handed counterpart (*43*). Alternatively, the two states may differ in their spatial delocalization on the helical lattice, such that the left-handed state overlaps more strongly with the lower-energy exciton states that receive its population and relaxes faster even at comparable spectral density (*41*). Energy-resolved excitation and exciton-vibration calculations on a constrained Hamiltonian could distinguish these. Finally, the relaxation rates are the same within error for the inner and outer nanotubes despite differences in their molecular packing (*20*), which places a strong constraint on any model decoding the electronic structure and dynamics.

Our results show that the helicity of light can steer excitations onto distinct relaxation pathways on ultrafast timescales, establishing supramolecular chirality as a handle for routing excited-state energy. Such helicity-selective control could underpin enantioselective photochemistry (*6*), chiroptical information processing (*45*), and the optical initialization of spin states in chiral molecular platforms for quantum information science (*7*).

**Acknowledgements**

This work was completed in part with resources provided by the University of Chicago Biophysics core facility (RRID: SCR_017915) and the University of Chicago Advanced Electron Microscopy Core Facility (RRID:SCR_019198).

The authors acknowledge financial support from the National Science Foundation through Quantum sensing for Biophysics and Bioengineering (QuBBE) Quantum Leap Challenge Institute (QLCI) (NSF OMA-2121044), through grant number 1900359.

CCB acknowledges the support from the Department of Defense National Defense Science and Engineering Graduate (DoD NDSEG) Fellowship program.

IG thanks the Department of Chemistry, University of Chicago, for the Eugene Olshansky Memorial Merit Scholarship for funding.

**Author contributions:**

Conceptualization: IG, KVM

Synthesis and Characterization: KVM, CCB, IG, GSE

Nonlinear Spectroscopy: IG, KVM, CCB, GSE

Funding and supervision: GSE

Writing – original draft: IG, GSE

Writing – review and editing: IG, KVM, CCB, GSE

**Competing interests:** The authors declare that they have no competing interests.

**Data, code and materials availability:** All data supporting the findings of this study are detailed in the paper and the Supplementary Information. Raw data and analysis scripts can be made available upon reasonable request.

**Supplementary Materials:** Supplementary Information file

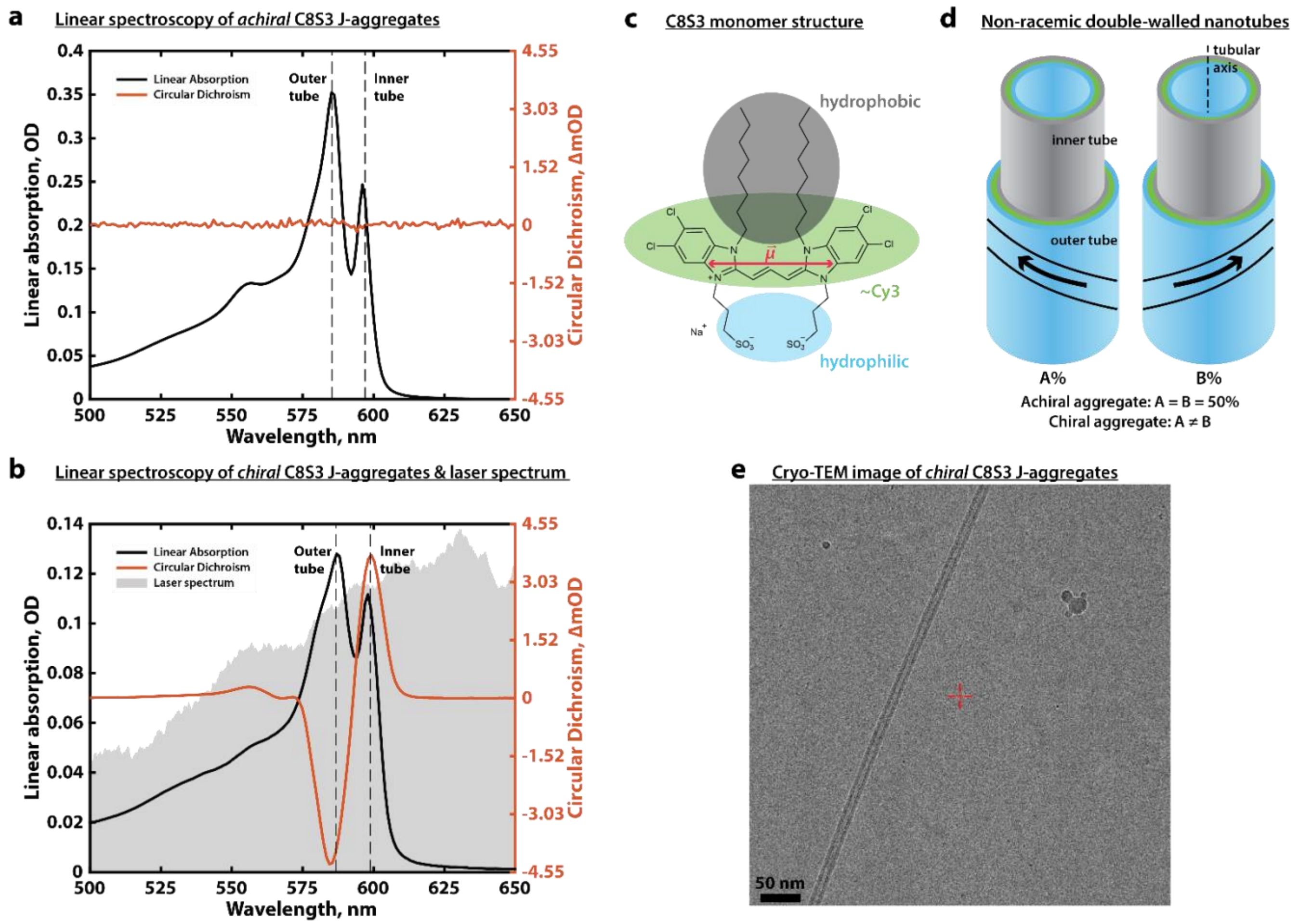

**Figure 1: C8S3 chiral aggregate. (a), (b)** Linear absorption and circular dichroism spectra of the racemic and chiral C8S3 J-aggregates, along with the laser spectrum of light pulses used in ultrafast spectroscopic measurements. The enantiomeric bias to the aggregate is provided by L-histidine. Vertical dashed lines denote peak positions of the inner and outer tube longitudinal excitations. **(c)** Molecular structure of the C8S3 monomer. C8S3 has a cyanine moiety structurally similar to Cy3, a hydrophilic chain ending with sulphate groups, and a hydrophobic group of hydrocarbon chains, making the molecule amphiphilic. The lowest energy electric transition dipole is shown as $\vec{\mu}$. **(d)** Schematic of the non-racemic mixture of double-walled helical nanotubes with colors matching the monomer moieties in **(c)**. **(e)** Cryogenic transmission electron microscopy image of chiral C8S3 J-aggregates showing double-walled nanotubes.

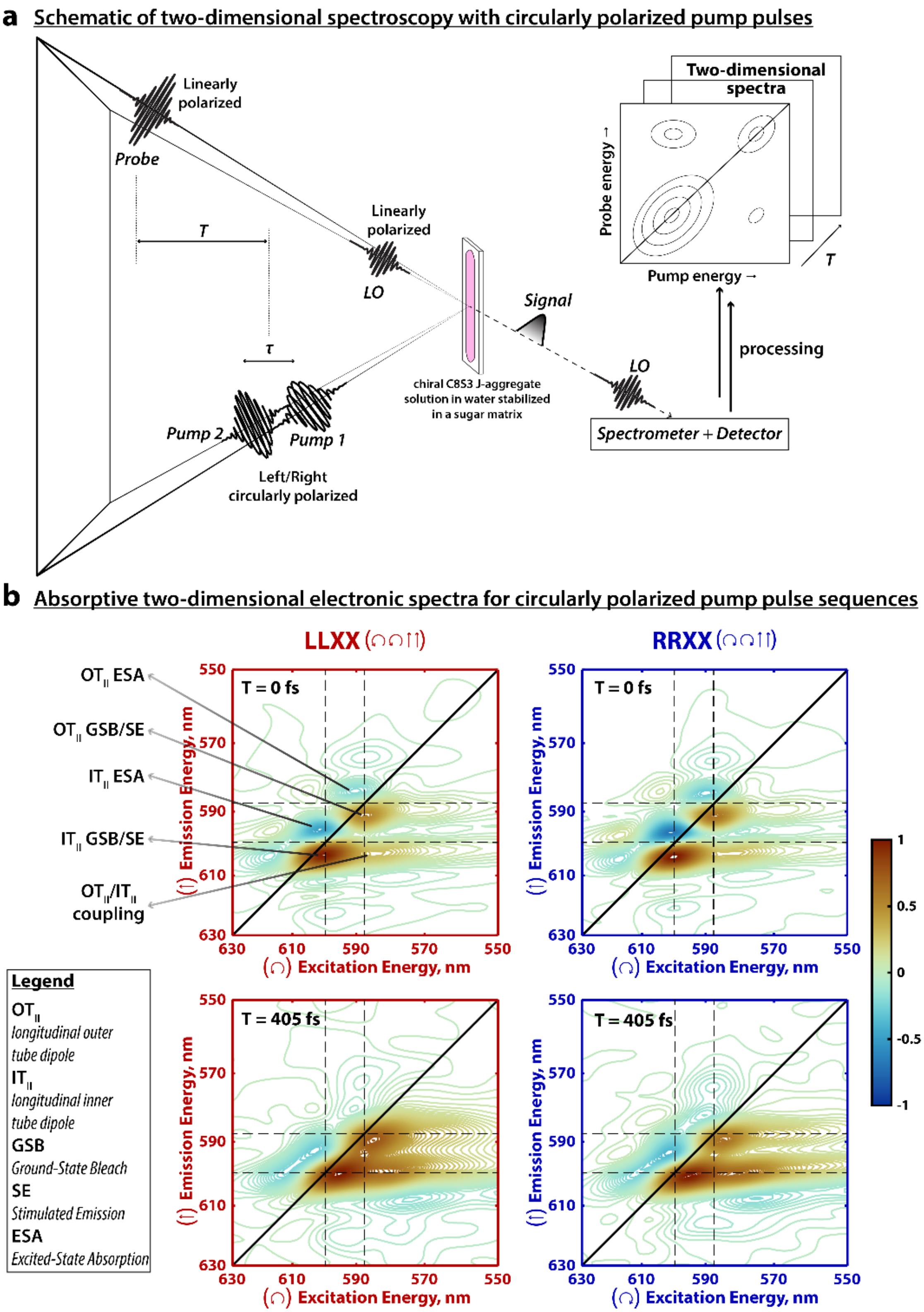

**Figure 2: Polarization-resolved two-dimensional electronic spectroscopy of chiral C8S3 aggregates. (a)** Schematic of the two-dimensional electronic spectroscopy (2DES) technique. Circularly polarized pump pulses

(left or right) excite the nanotubes, and linearly polarized probe and local oscillator (LO) pulses detect the signal. This measurement resolves electronic excited state population dynamics in time as a function of excitation and detection energies. **(b)** 2DES spectra of the $\langle LLXX \rangle$ (left) and $\langle RRXX \rangle$ (right) pulse sequences at waiting times (T) of 0 fs (top) and 405 fs (bottom). Positive features correspond to ground state bleach and stimulated emission signals, and negative features correspond to excited state absorption (*36*), as labeled in the top-left spectrum for both nanotubes. Dashed vertical and horizontal lines denote the wavelength positions of the inner tube and outer tube, as in **Fig. 1B**. Each frame is individually normalized, and 80 contours are plotted per frame.

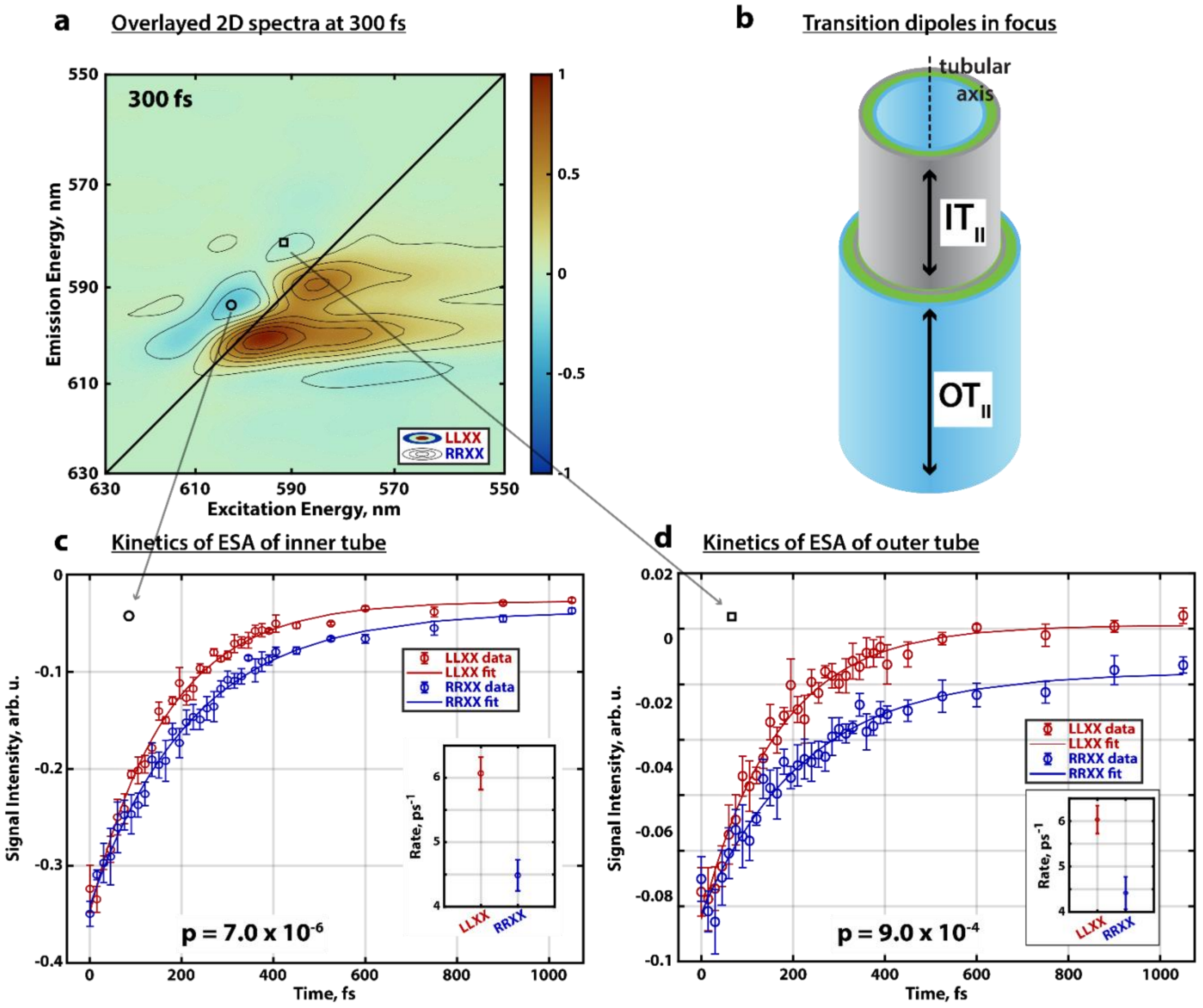

**Figure 3: Evidence of helicity-dependent exciton dynamics. (a)** Overlaid 2DES absorptive spectra at T = 300 fs for the $\langle LLXX \rangle$ (filled colored contours) and $\langle RRXX \rangle$ (black contour lines) pulse sequences. **(b)** Schematic of the J-aggregate nanotubes showing transition dipoles relevant to this figure. **(c, d)** Time traces for the $\langle LLXX \rangle$ and $\langle RRXX \rangle$ sequences at **(c)** $\lambda_{ex}$=602.7 nm, $\lambda_{em}$=593.0 nm and **(d)** $\lambda_{ex}$=591.6 nm, $\lambda_{em}$=581.1 nm. Error bars indicate standard error of the mean (SEM) obtained from four repeat of each pulse sequence. Insets in (**c**) and (**d**) show the fitted rates (mean ± SEM), obtained by fitting each repeat separately, the P-values from F-tests comparing the $\langle LLXX \rangle$ and $\langle RRXX \rangle$ traces.

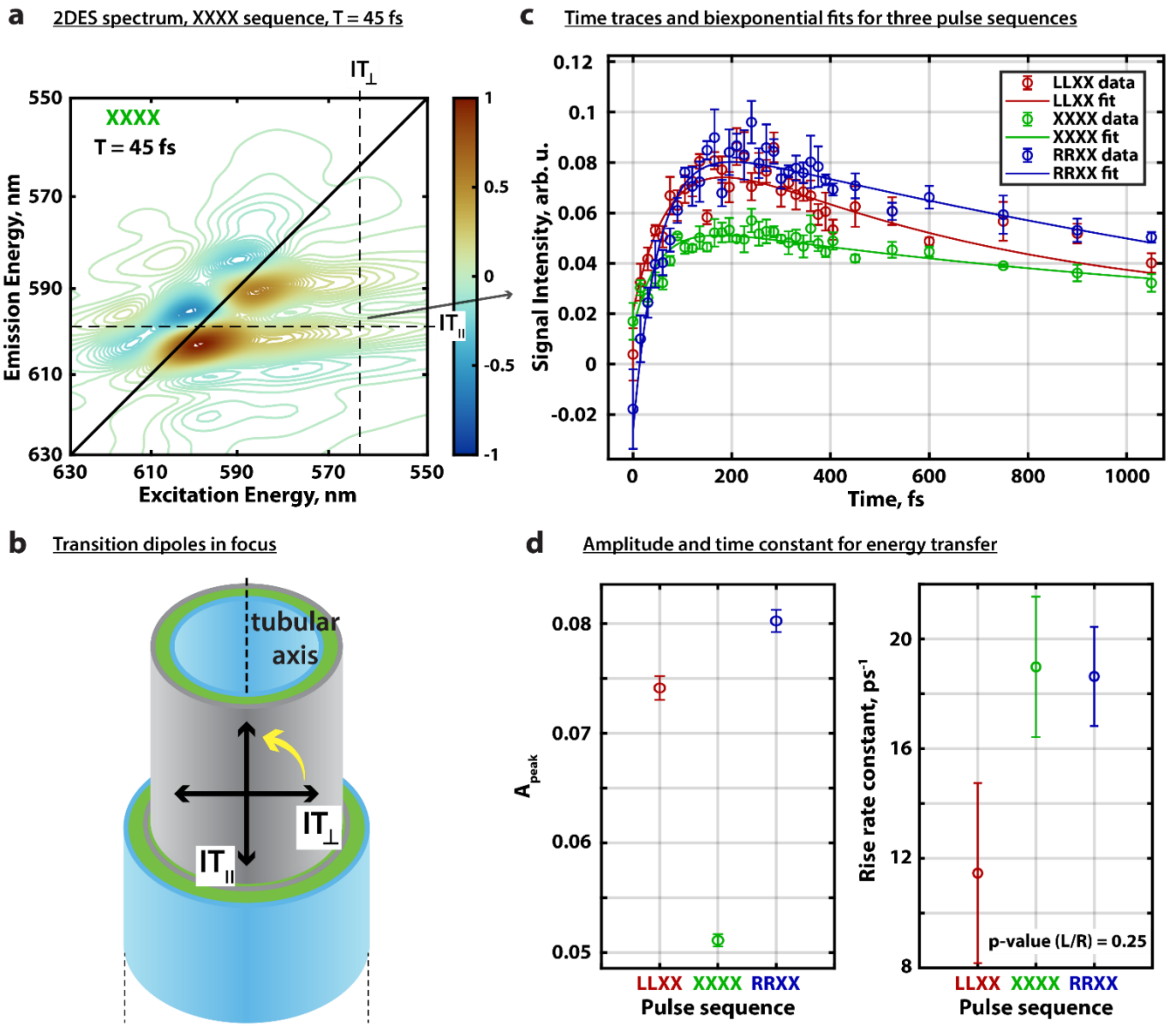

**Figure 4: Helicity-independent transfer from transverse excitons. (a)** 2DES absorptive spectrum at T = 45 fs for the $\langle XXXX \rangle$ pulse sequence. Dashed lines intersect at $\lambda_{ex}$=564.4 nm, $\lambda_{em}$=599.4 nm, marking the below-diagonal cross peak corresponding to energy transfer from the transverse to the longitudinal excited states. **(b)** Schematic of the J-aggregate nanotube showing the transverse and longitudinal transition dipoles. The yellow arrow indicates energy transfer from the transverse to the longitudinal dipole. **(c)** Time traces for $\langle LLXX \rangle$, $\langle XXXX \rangle$, and $\langle RRXX \rangle$

pulse sequences at $\lambda_{ex}$=564.6 nm, $\lambda_{em}$=597.4 nm, with biexponential fits. Error bars are SEM from four repeats of each sequence. **(d)** Amplitude and rise rate of the cross-peak signal obtained from fits in **c**, shown as mean ± SEM. Rise rates show no significant different among the three sequences (F-test, P = $0.25$ quoted for $k_r$ for the $\langle LLXX \rangle$ and $\langle RRXX \rangle$ sequences).

Supplementary Information

# Handed Ultrafast Dynamics in Chiral Cyanine Nanotubes

Indranil Ghosh[§,†], Keerthi Vaasan Mani[§,†], Caitlin C. Bellora[§], Gregory S. Engel[§,*]

[§]Department of Chemistry, Institute for Biophysical Dynamics, James Franck Institute, and Pritzker School of Molecular Engineering at The University of Chicago, Chicago, IL 60637, USA

[†]Contributed equally

*Corresponding author: gsengel@uchicago.edu

## 1. Sample Preparation

The amphiphilic cyanine dye derivative 3,3'-bis(2-sulfopropyl)-5,5',6,6'-tetrachloro-1,1'-dioctylbenzimidacarbocyanine (C8S3, molecular weight 902.8 g.mol$^{-1}$) was obtained as a sodium salt (FEW Chemicals). 0.64 mg of monomer dye was weighed and added to 464 µL methanol to make a 1.46 mM stock solution. Upon addition of water, C8S3 forms intrinsically helical J-aggregates (double-walled nanotubes), but it is a racemic ensemble of equal P and M helices, thereby resulting in no net CD signal.(*1*) To induce an enantiomeric bias of one form of helical aggregate, we prepare 467 µL of 15.72 mM solution of L-Histidine in milliQ water, to which we slowly add 121 µL of the stock solution. The resulting solution containing water, methanol and L-Histidine has a final monomer concentration of 0.3 mM. For ease of nomenclature, we call the aggregate sample with the enantiomeric bias as "chiral aggregates", but we are only creating a non-racemic mixture of helical nanotubes which would otherwise be racemic without a chirality-inducing agent. Two parts (by volume) of the aggregate solution was slowly added into one part of sugar solution (1:1 w/w, Trehalose + Sucrose in milliQ water)(*2*) and the mixture was drop-casted uniformly onto quartz glass to make a thin film. The sugar film was heated with a heat block for about 10 minutes at 40 °C.

## 2. Calculation of Chiral Dissymmetry Factor

The chiral dissymmetry factor is defined as follows.

$$g_{abs} = \Delta\varepsilon(\lambda) / \varepsilon(\lambda) \qquad \text{Equation S1}$$

Here, $\Delta\varepsilon = \varepsilon_L - \varepsilon_R$ where $\varepsilon_L$ and $\varepsilon_R$ denote the molar extinction coefficients for left and right circularly polarized light respectively, and $\varepsilon$ is the molar extinction coefficient for unpolarized light.

The conversion factors for $\varepsilon$ from absorbance ($A$) and $\Delta\varepsilon$ from ellipticity ($\theta$) are $\Delta\varepsilon = \frac{\theta(mdeg)\times 10^{-3}}{32890\times c(M)\times l(cm)}$, and $\varepsilon = \frac{A}{c.l}$, where $c$ and $l$ are molar concentration and cuvette path length respectively. Using these conversion factors, we obtain a relation between $g_{abs}$, $A$ and $\theta$. The values of $A$ and $\theta$ are as shown in **Figures S2a** and **S5a**.

$$g_{abs} = \frac{\theta(\lambda)}{32980\times A(\lambda)} \qquad \text{Equation S2}$$

At the inner tube's spectral position (599 nm), we have $g_{abs,599\ nm} = 0.036$, and at the outer tube's spectral position $g_{abs,587\ nm} = 0.036$.

## 3. Sample Characterization and Discussion

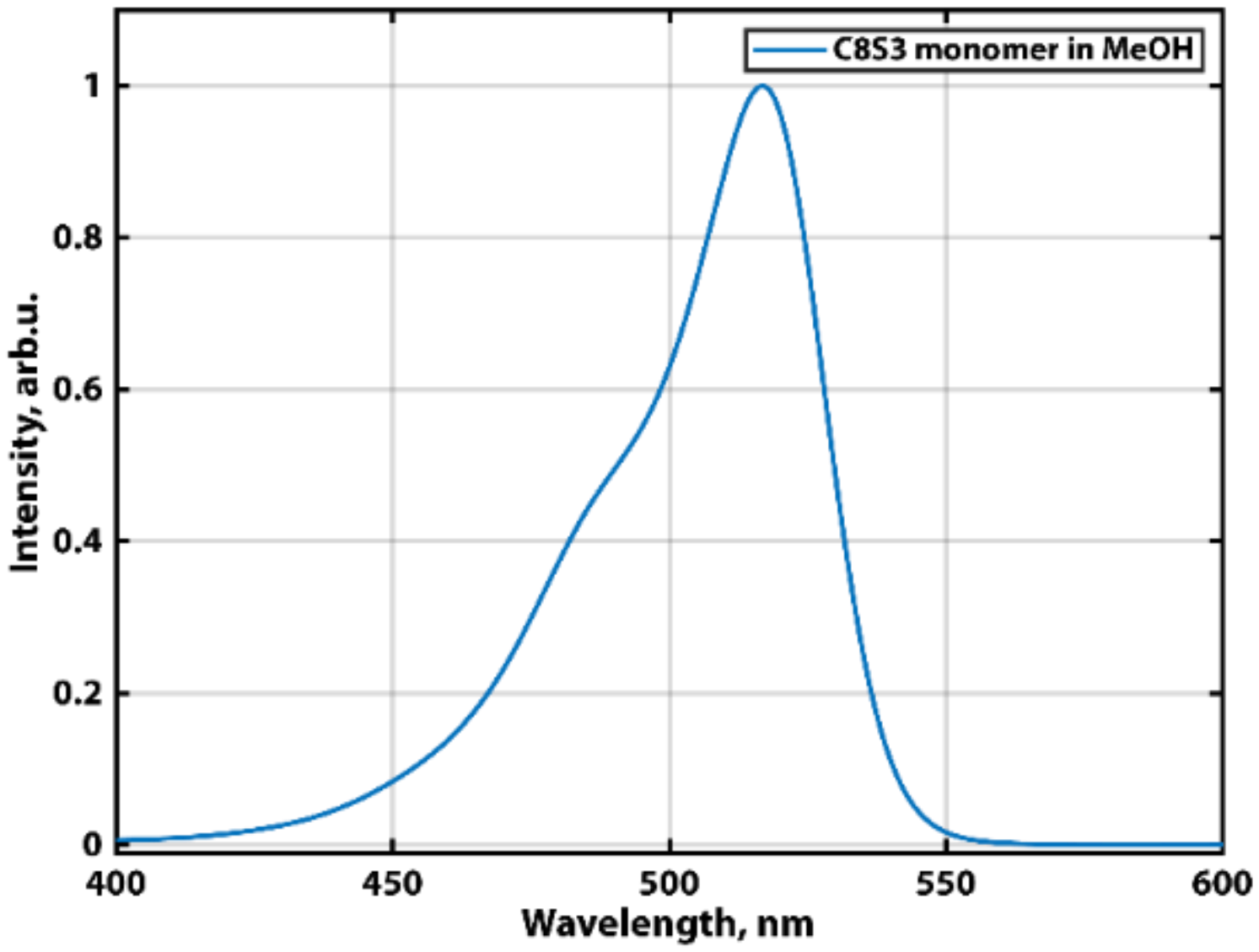

**Figure S1**: Absorption of the C8S3 monomer dissolved in anhydrous methanol.

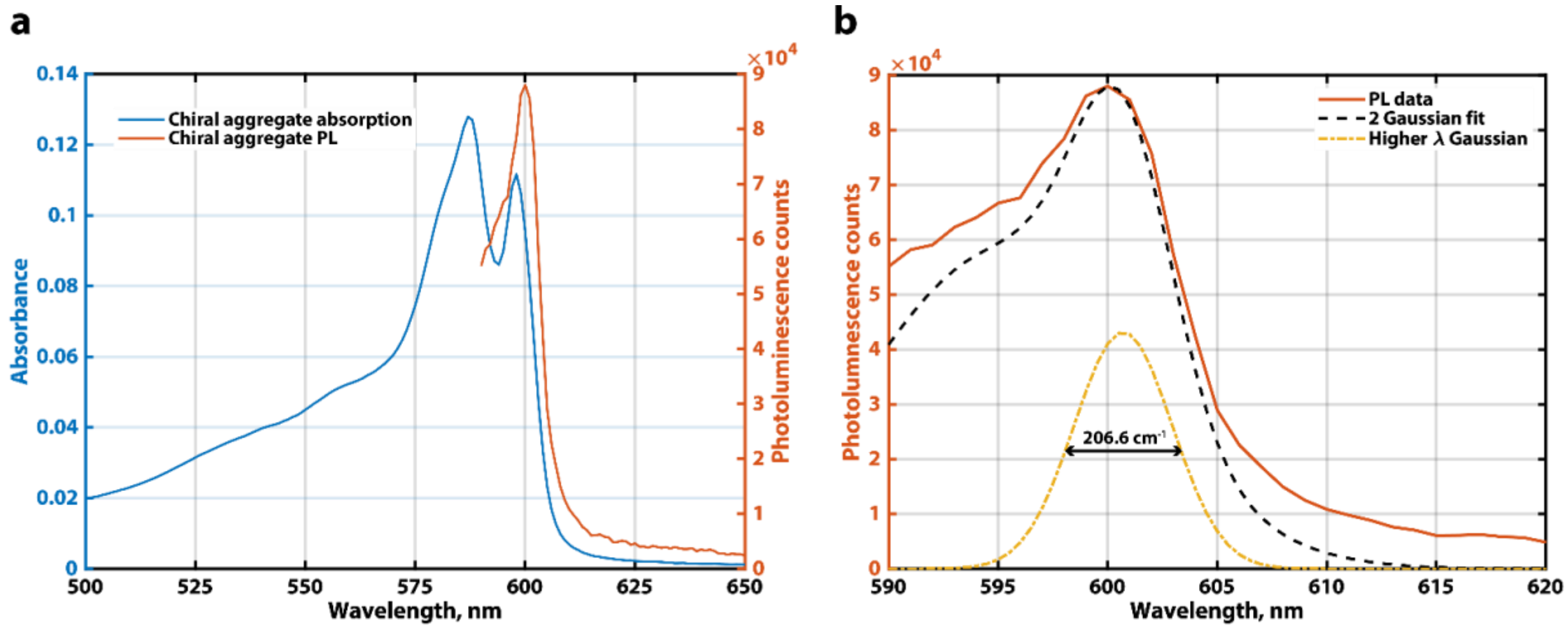

**Figure S2**: **(a)** Absorption and photoluminescence of chiral C8S3 aggregates used in 2DES measurements after excitation with unpolarized light. The photoluminescence was obtained after exciting at 580 nm. **(b)** Sum of two Gaussians fitted to the photoluminescence (PL) data. The lower energy (higher λ) Gaussian is centered at 600.7 nm and has full-width at half maximum of 206.6 $cm^{-1}$. The Stokes shift between the inner tube longitudinal transition dipole centered at 598.7 nm (see **Table S1**) and 600.7 nm (see **Figure S2**) is 55.6 $cm^{-1}$.

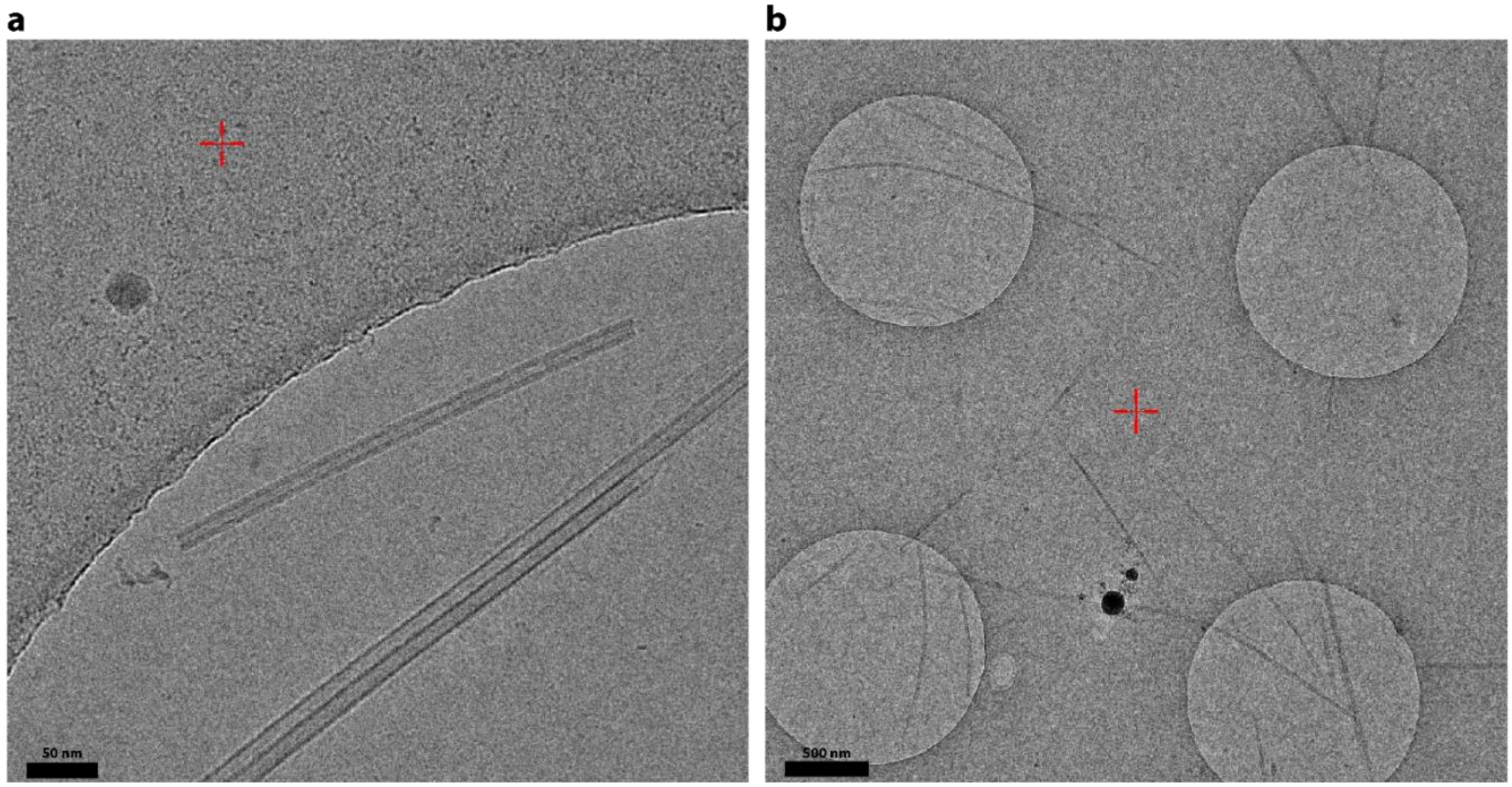

**Figure S3**: Additional Cryo-TEM images of the chiral aggregate.

**Discussion on electronic structure of double-walled nanotubes**

The smallest unit of the nanotubular morphology is a dimer of two C8S3 monomers, giving rise to two electronic states from the in-phase and out-of-phase overlap of the monomeric excited state wavefunctions (*1*). Creating a nanotube out of this dimer unit creates four excitonic states: two polarized parallel to the tubular axis (longitudinal excitons) and two perpendicular to the tubular axis (transverse excitons). Therefore, two concentric nanotubes correspond to eight excitonic states (*3*), out of which the two most intense absorption peaks corresponding to the lower energy inner and outer tube longitudinal excitons absorb at 599 nm and 587 nm, respectively. Another possible morphology is bundles of single-tube aggregates (*4*). This results in a red-shifted sharp peak at ~602 nm corresponding to the longitudinal inner tube exciton and a broad peak centered at 575 nm that signifies bundling. Caram and co-workers' high resolution cryo-TEM study showed that the "bricklayer" stacking (*5*) of the monomers is helical for a given tube.

We note that the chiral aggregate lineshapes are relatively broad (see **Figure S2, S4**), which can be from coupling to intra- and intermolecular vibrations and static disorder primarily induced by L-histidine's interaction with the nanotube. A 55 $cm^{-1}$ Stokes-shifted fluorescence spectrum for the inner tube also attests to weak coupling to low-frequency bath vibrations. To rule out bundling in our chiral aggregate sample, we fit five Gaussians to the linear absorption spectra of both the racemic and chiral aggregates and show the first three fitted Gaussians in descending order of wavelength in **Figure S4**. The Gaussian fits show largely unchanged peak positions between the racemic and chiral samples and, more importantly, no broad peak at or near 575 nm (also see **Table S1**).

Intrinsic chiroptical activity originates from the magnetic transition dipole of a chromophore. Fundamentally, the differential absorption between left and right circularly polarized light (ECD signal) is proportional to the rotational strength given by $Im(\vec{\mu} \cdot \vec{m})$, where $\vec{\mu}$ and $\vec{m}$ are the electric and magnetic transition dipoles, respectively (*6*, *7*). In molecular aggregates, a rotation of orientation between displaced electronic transition dipoles also gives rise to a ECD signal and is called excitonic chirality, which goes beyond the point dipole approximation. In this case, the cross product between the two electronic dipoles replaces the magnetic transition dipole in the expression for rotational strength, as shown by Painelli and co-workers (*7*). A recent hierarchical model to describe a chiral bundled aggregate of C8O3 used the "herringbone" dipolar arrangement and required the C8O3 monomer to be non-planar (implying non-zero dot product of electric and magnetic transition dipoles) to compute the circular dichroism spectrum of the aggregate (*8*). However, high-resolution linear dichroism microscopy and cryo-TEM studies provide strong evidence for a "bricklayer" arrangement of dipoles (*9*, *1*, *5*). Although we posit that the ECD spectral features for our chiral aggregate primarily originate from excitonic chirality, a definitive structural model to simulate the spectra is beyond the scope of this report. A fit of the ECD spectrum with four Gaussians is given in **Figure S5** and **Table S2**. Incorporating the structural role of excess L-histidine into the nanotubular arrangement may be critical in determining the three-dimensional arrangement of dipoles and consequently simulating the ECD spectrum.

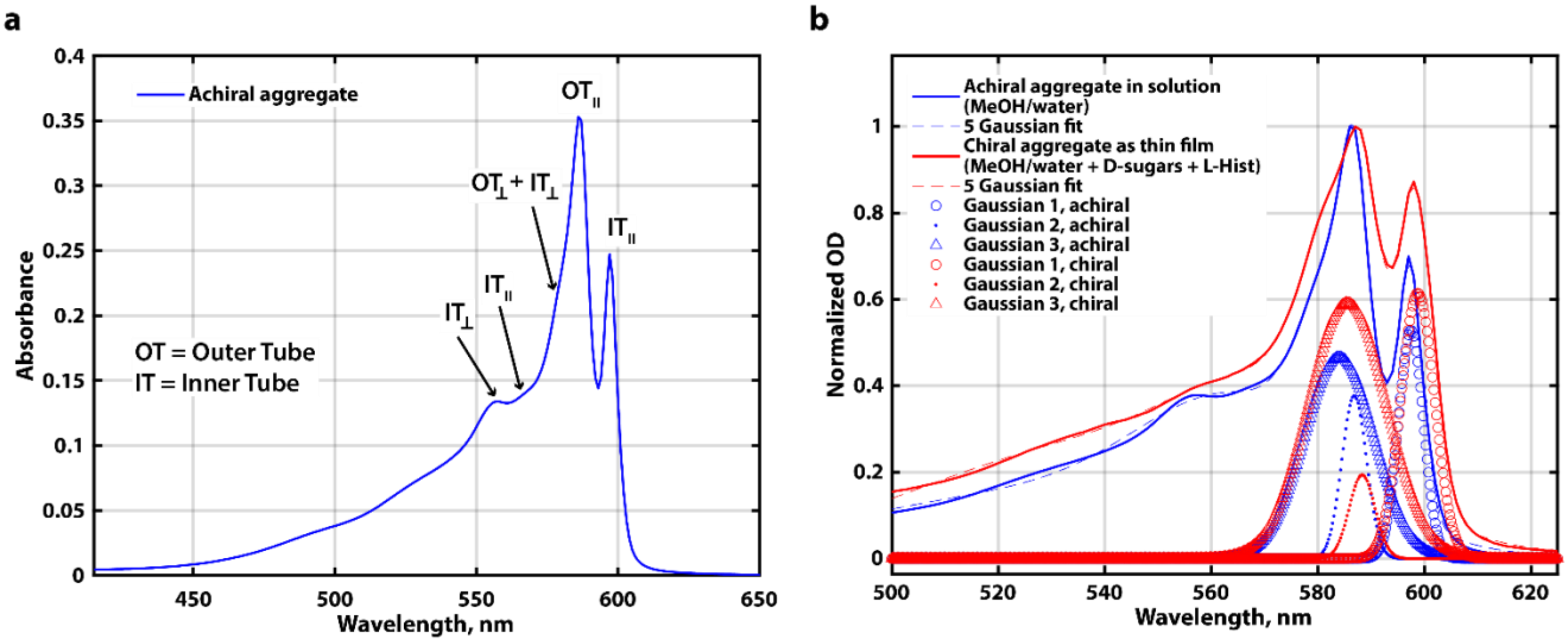

**Figure S4**: **(a)** Normalized absorption spectra of the achiral aggregate in methanol/water. The suffixes of the OT and IT labels stand for transition dipoles aligned parallel and perpendicular to the tube axis. Parallel and perpendicular symbols in the label suffixes denote longitudinal and transverse transition dipole alignments with respect to the tubular axis. **(b)** Comparison of the absorption spectra of the achiral aggregate (same as in **a**) and the chiral aggregate that was used for 2DES measurements with a sum of five Gaussians fitted to both spectra. The first three Gaussian fits in order of descending center-wavelength values are plotted for both spectra.

**Table S1:** Fit parameters of the five Gaussians as shown in Figure S4.

| Gaussian parameters | Gaussian 1 | | Gaussian 2 | | Gaussian 3 | | Gaussian 4 | | Gaussian 5 | |
|---|---|---|---|---|---|---|---|---|---|---|
| | Achiral | Chiral | Achiral | Chiral | Achiral | Chiral | Achiral | Chiral | Achiral | Chiral |
| Relative amplitude | 0.53 | 0.61 | 0.38 | 0.19 | 0.47 | 0.59 | 0.29 | 0.22 | 0.15 | 0.27 |
| Center wavelength (nm) | 597.5 | 598.7 | 586.8 | 588.3 | 584 | 585.5 | 565.7 | 570.5 | 528 | 540.9 |
| Standard deviation (nm) | 3.3 | 4.5 | 3.3 | 3.4 | 9.1 | 10.2 | 26.2 | 22.8 | 52.8 | 50.5 |

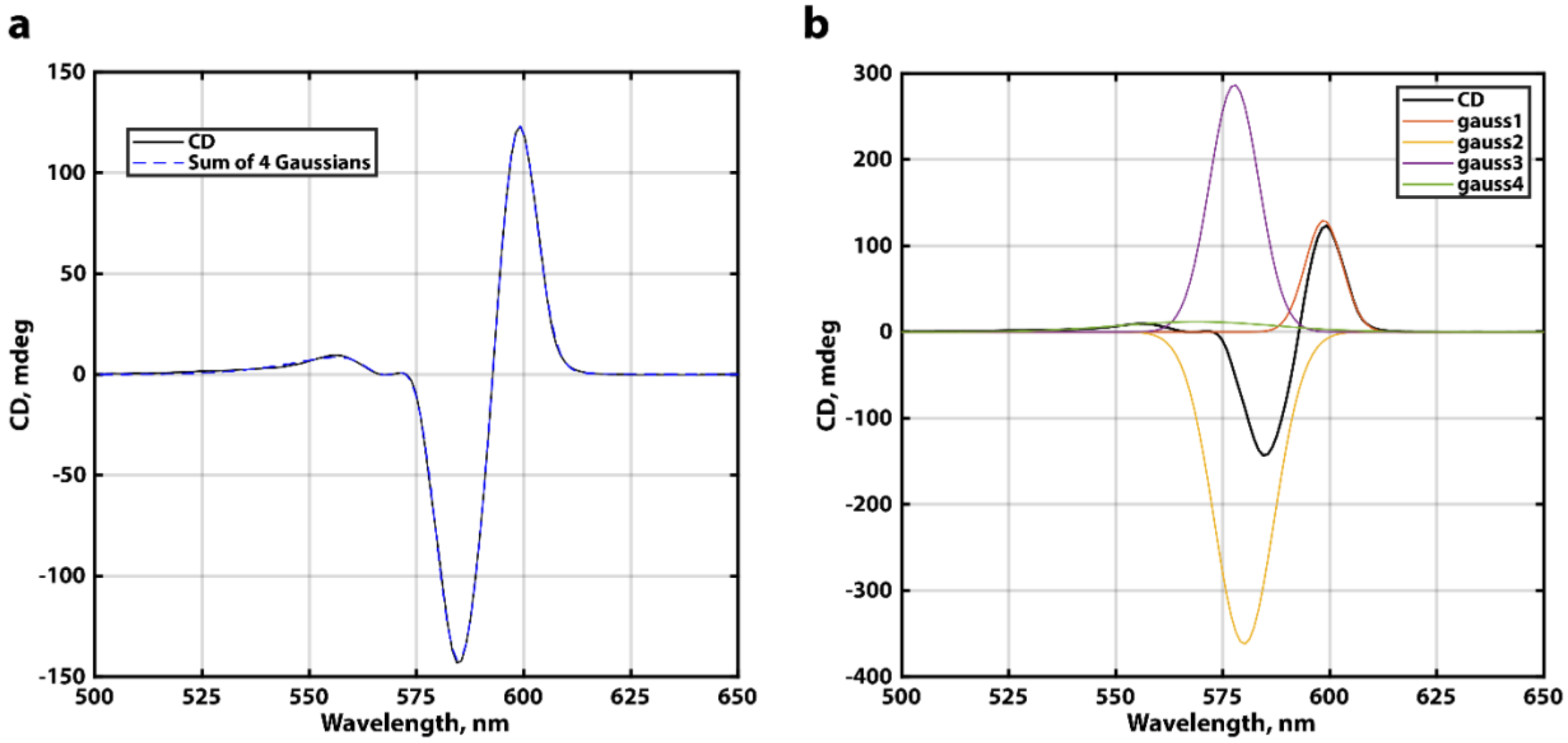

**Figure S5**: (**a**) CD spectrum of the chiral aggregate superimposed with the fitted curve which is the sum of four Gaussians. (**b**) Individual Gaussians plotted along with the experimental CD spectrum.

**Table S2:** Fit parameters of the four Gaussians as shown in Figure S5.

| Gaussian parameters | Gaussian 1 | Gaussian 2 | Gaussian 3 | Gaussian 4 |
|---|---|---|---|---|
| Relative amplitude | 129.3 | -362 | 286.8 | 11.8 |
| Center wavelength (nm) | 598.6 | 580 | 577.7 | 568.8 |
| Standard deviation (nm) | 6.14 | 9.86 | 8.17 | 26.07 |

## 4. Reasoning Behind Conducting *LLXX* and *RRXX* pulse sequences

It is convenient to convert a pulse sequence that contains both circular and linearly polarized pulses into one basis. Since we usually work with the electric dipole approximation, we use linearly polarized pulse sequences. Therefore, we convert our mixed circular/linear sequence into a linear pulse sequence. Using the IUPAC convention of circularly polarized light, $L = X + iY$, and $R = X - iY$, where $L$ and $R$ are left and right circularly polarized vectors constituting the circular basis, while $X$ and $Y$ constitute the linear basis. Without loss of generality, the phase matching direction for a rephasing pathway is $-\boldsymbol{k_1} + \boldsymbol{k}_2 + \boldsymbol{k}_3$; therefore, the polarization combination is going to be $\{\varepsilon_1^*, \varepsilon_2, \varepsilon_3, \varepsilon_{LO}^*\}$ where $\varepsilon_i$ indicates the polarization of the $i$th pulse and * denotes complex conjugation. This means if $\varepsilon_1 = X + iY$, then $\varepsilon_1^* = X - iY$. Similarly, the non-rephasing pathway is phase-matched to $+\boldsymbol{k_1} - \boldsymbol{k}_2 + \boldsymbol{k}_3$, giving the polarization combination of $\{\varepsilon_1, \varepsilon_2^*, \varepsilon_3, \varepsilon_{LO}^*\}$. Using this machinery, we can write the following relations.

$$\langle LLXX\rangle_{rephasing} = \frac{1}{2}[\langle XXXX\rangle + \langle YYXX\rangle + i(\langle XYXX\rangle - \langle YXXX\rangle)]$$

$$\langle LLXX\rangle_{non-rephasing} = \frac{1}{2}[\langle XXXX\rangle + \langle YYXX\rangle - i(\langle XYXX\rangle - \langle YXXX\rangle)]$$

$$\langle RRXX\rangle_{rephasing} = \frac{1}{2}[\langle XXXX\rangle + \langle YYXX\rangle - i(\langle XYXX\rangle - \langle YXXX\rangle)]$$

$$\langle RRXX\rangle_{non-rephasing} = \frac{1}{2}[\langle XXXX\rangle + \langle YYXX\rangle + i(\langle XYXX\rangle - \langle YXXX\rangle)] \quad \text{Equation S3}$$

An odd pulse sequence is where an $X$ or $Y$ polarization appears odd number of times (third and fourth terms in **Equation S3**). Minhaeng Cho(*10*) has shown before than odd linearly polarized four-wave sequences are devoid of the electric dipolar response and therefore, the leading response terms in those pulse sequences come from electric quadrupolar and magnetic dipolar (chiral) response. We observe that only odd pulse sequences, *viz.*, $\langle XYXX\rangle$ and $\langle YXXX\rangle$ survive if we subtract $\langle LLXX\rangle$ from $\langle RRXX\rangle$. Therefore, any differences in dynamics between the $\langle LLXX\rangle$ and $\langle RRXX\rangle$ pulse sequences cannot come from electric dipolar response.

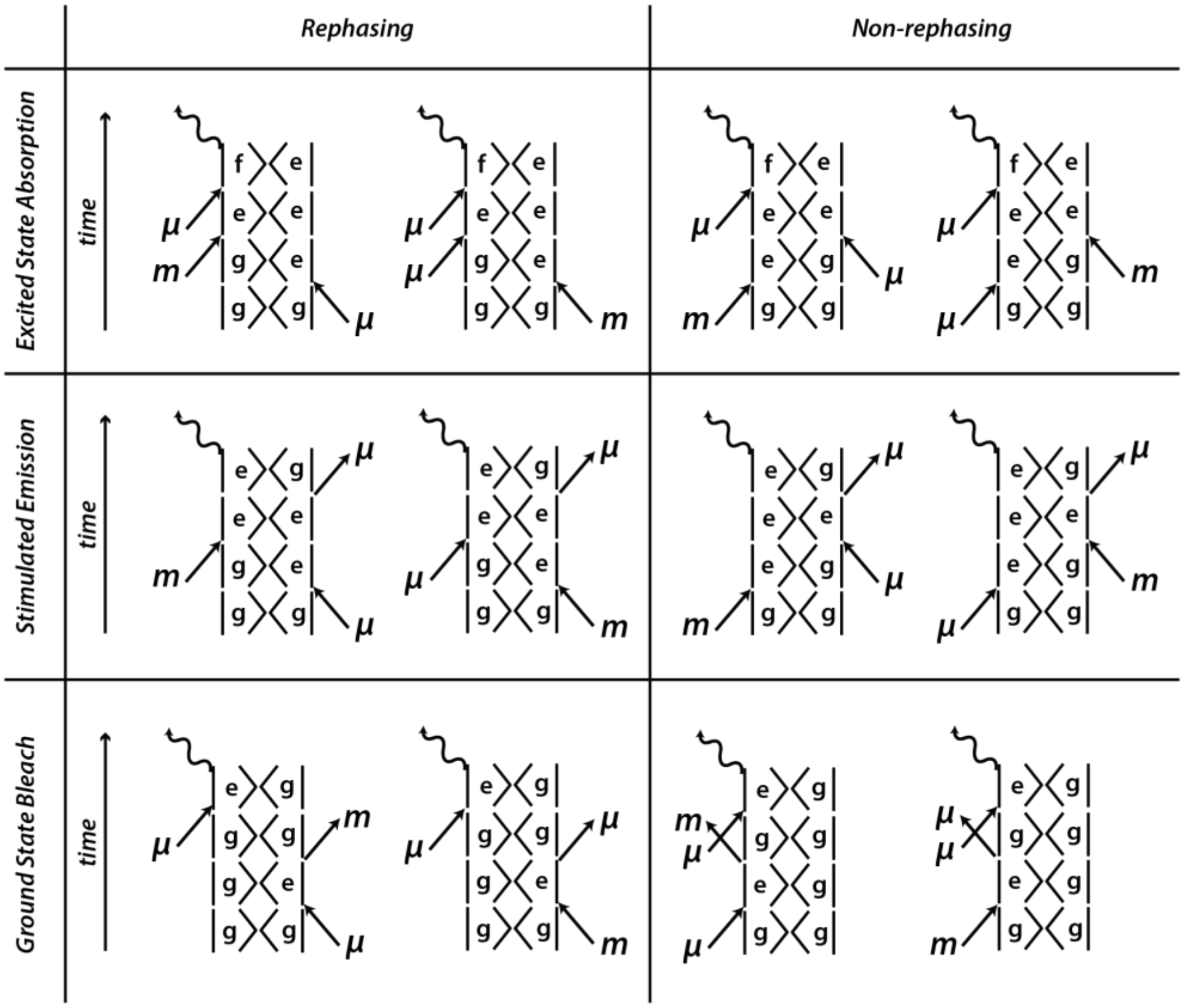

**Figure S6**: Third order nonlinear Feynman pathways that survive after subtracting signals from the $\langle LLXX\rangle$ and $\langle RRXX\rangle$ datasets. $|g\rangle$ = ground state, $|e\rangle$ = first excited state, $|f\rangle$ = higher excited state. Interaction with the electric quadrupole (and higher orders) and magnetic quadrupole (and higher orders) are ignored.

It is important to note that the signal contribution from the above pathways will flip sign in the subtracted spectra, because there is an overall negative sign in Equation S3. For example, a negative signal would correspond to an excited state absorption (ESA) pathway, but due to the negative sign, a positive signal in the subtracted spectra would correspond to ESA.

## 5. 2D Spectroscopy of Chiral Aggregates: Additional data

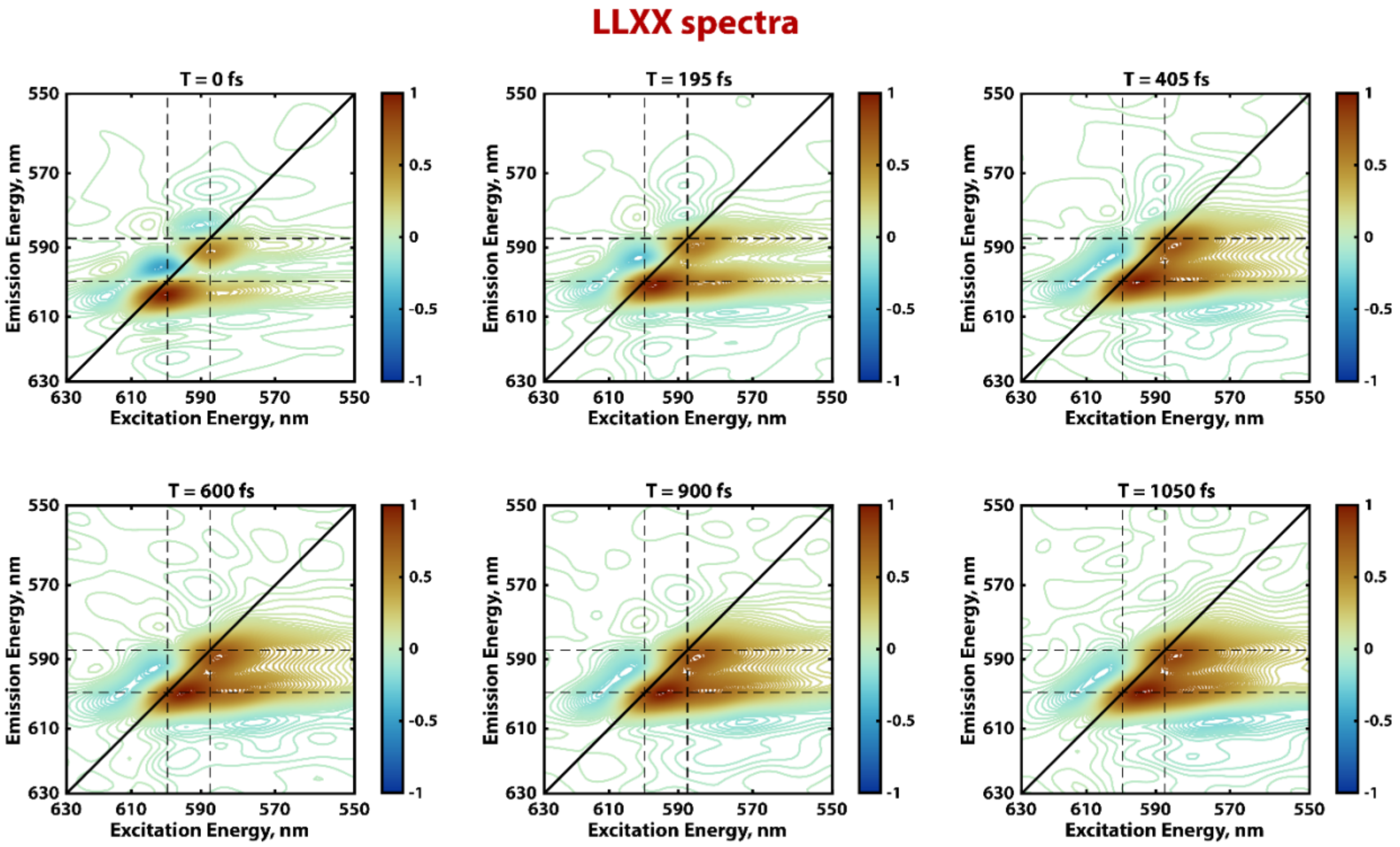

**Figure S7**: Chiral aggregate additional absorptive two-dimensional electronic spectra with the $\langle LLXX \rangle$ pulse sequence. All frames individually normalized.

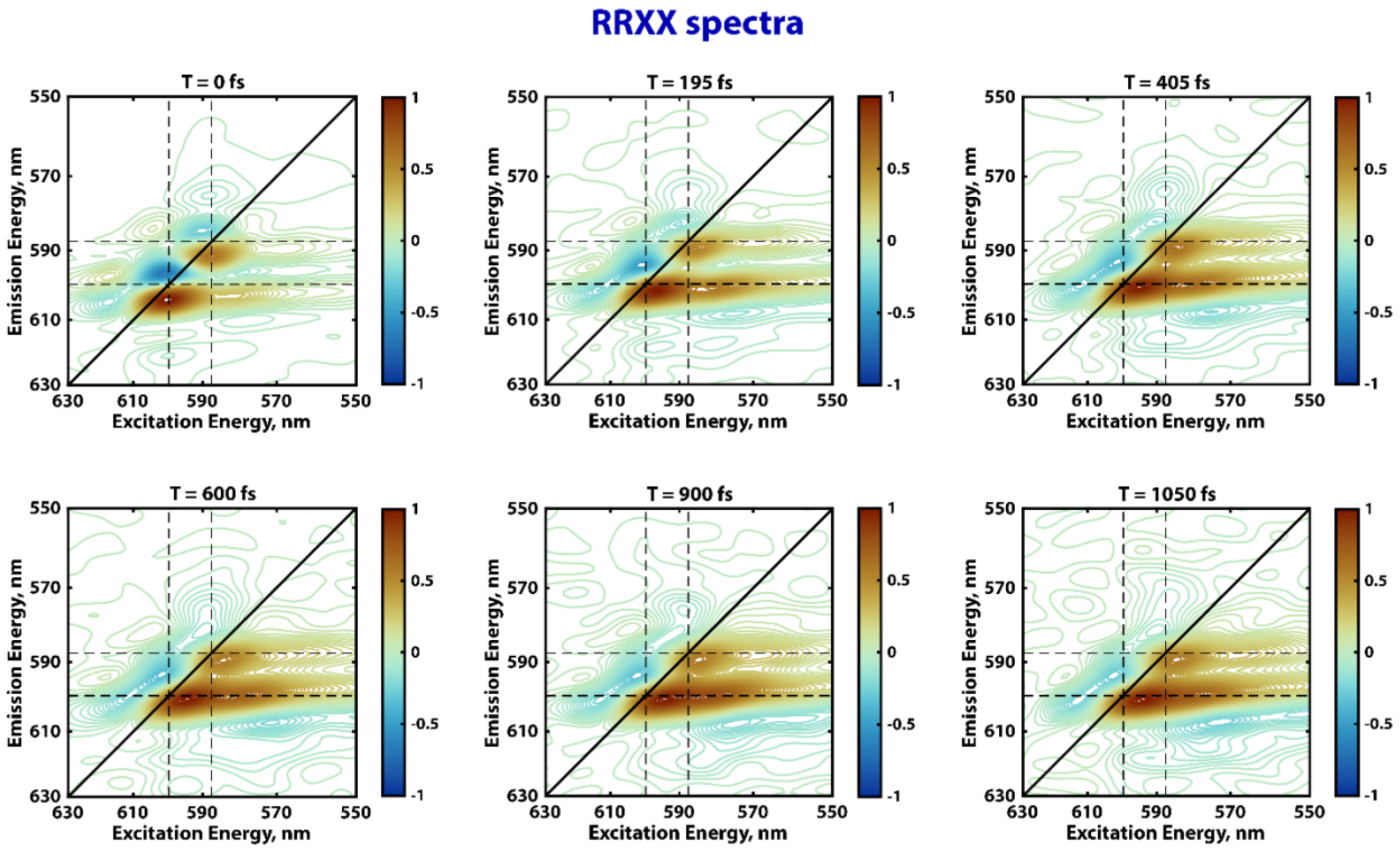

**Figure S8**: Chiral aggregate additional absorptive two-dimensional electronic spectra with the $\langle RRXX \rangle$ pulse sequence. All frames are individually normalized.

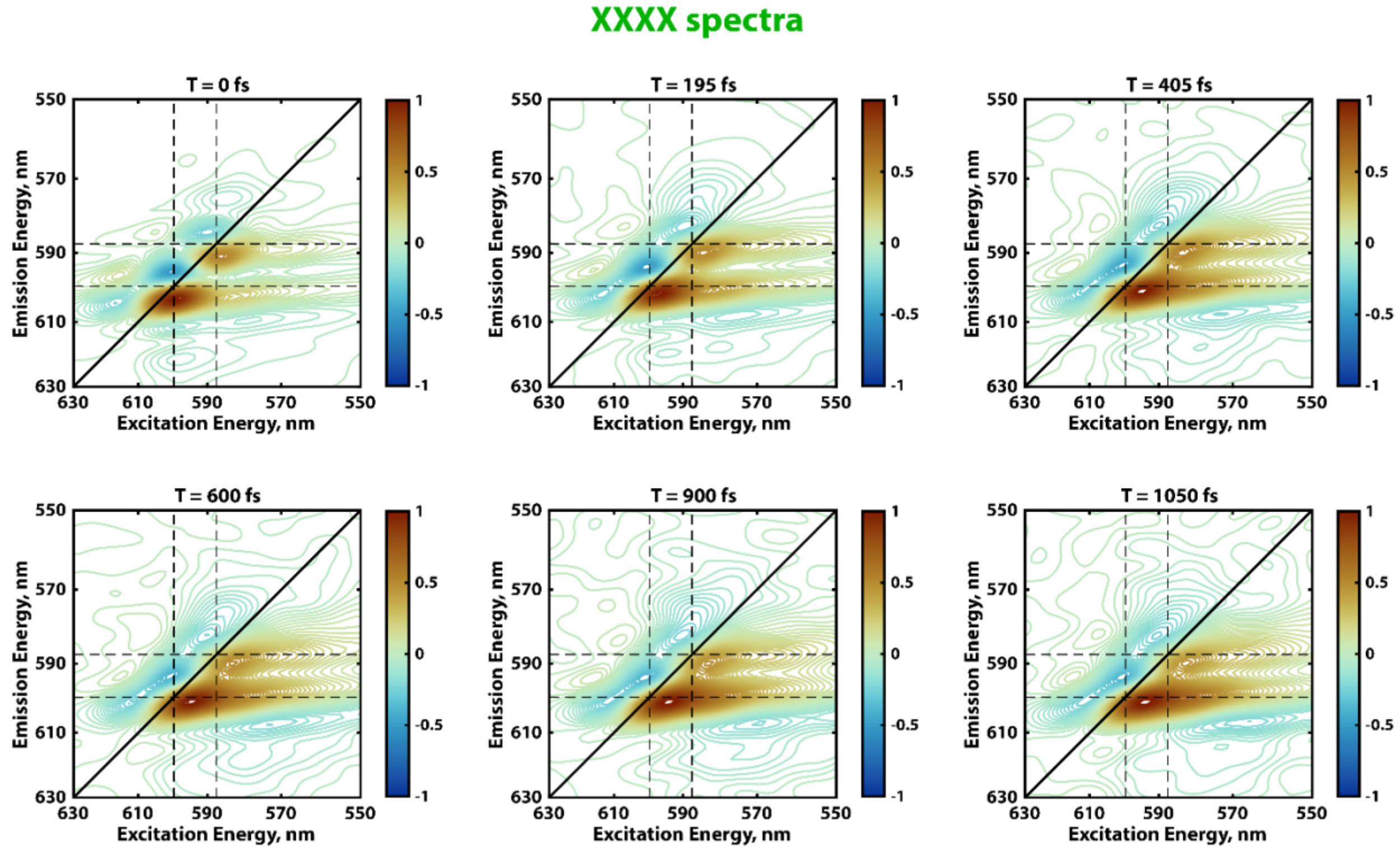

**Figure S9**: Chiral aggregate absorptive two-dimensional electronic spectra with the $\langle XXXX \rangle$ pulse sequence. All frames are individually normalized.

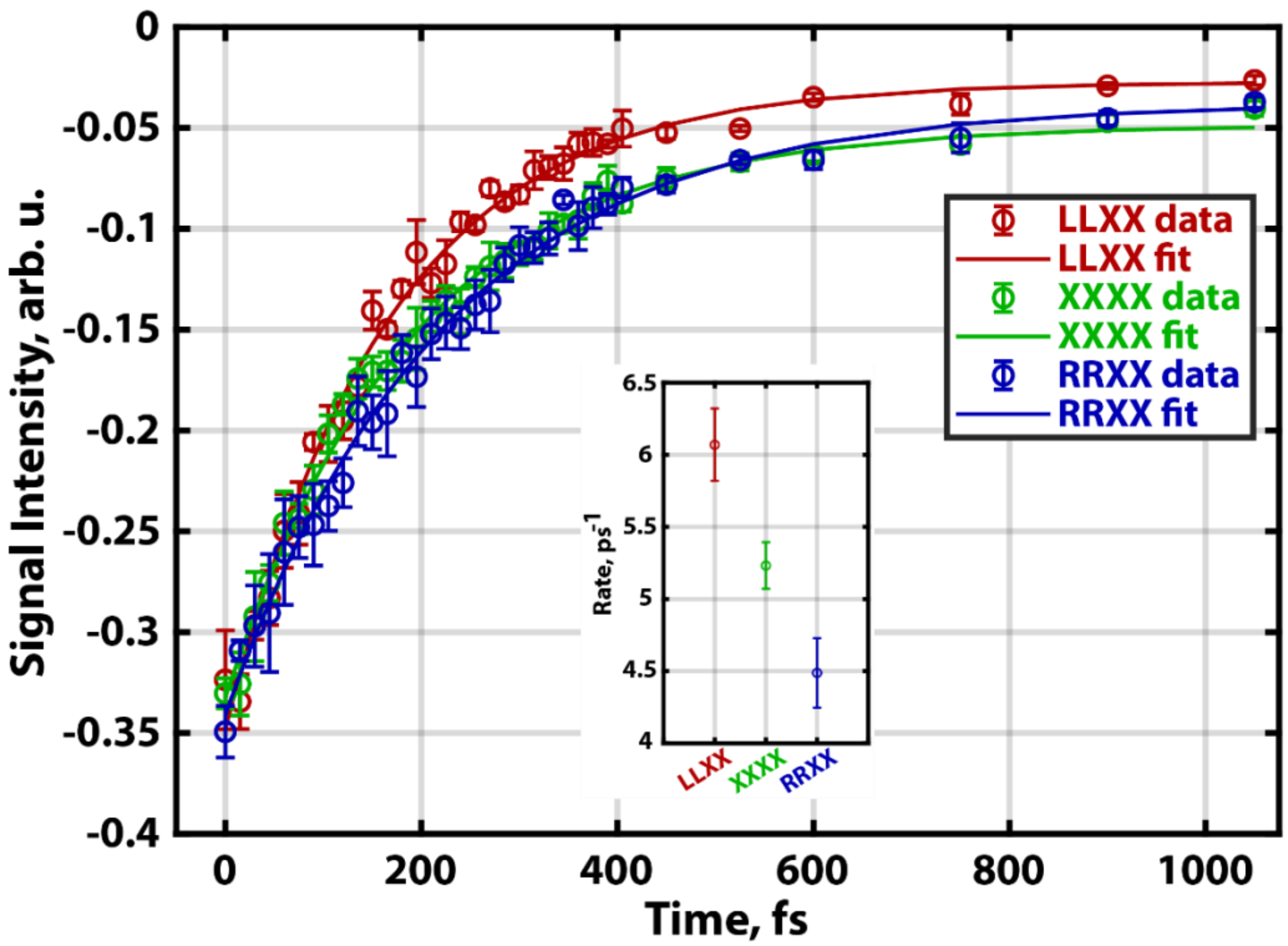

**Figure S10:** The same time traces as plotted in Figure 3c ($\lambda_{ex}$=602.7 nm, $\lambda_{em}$=593.0 nm) are shown here along with that of the all-parallel ($\langle XXXX \rangle$) pulse sequence.

**<u>Positive cross-peak between the inner and outer longitudinal excitons</u>**

Contrary to expectation (*11*), we do not directly observe excitation energy transfer between the outer and inner tube longitudinal excited states in our room temperature measurements, which would be manifested as a rise of the positive cross peak below the diagonal in waiting time. An analysis of this cross-peak signal is presented in **Figure S11.** We indeed observe a positive below-diagonal cross-peak at zero waiting time (see **Figure 2B**), indicating weak coupling between the inner and outer tubes, but that feature decays in waiting time with the same rate as the diagonal features indicating a ground-state Feynman pathway, while a rise in signal would indicate stimulated emission coming from excitation energy transfer from outer to inner tube longitudinal states (*11*).

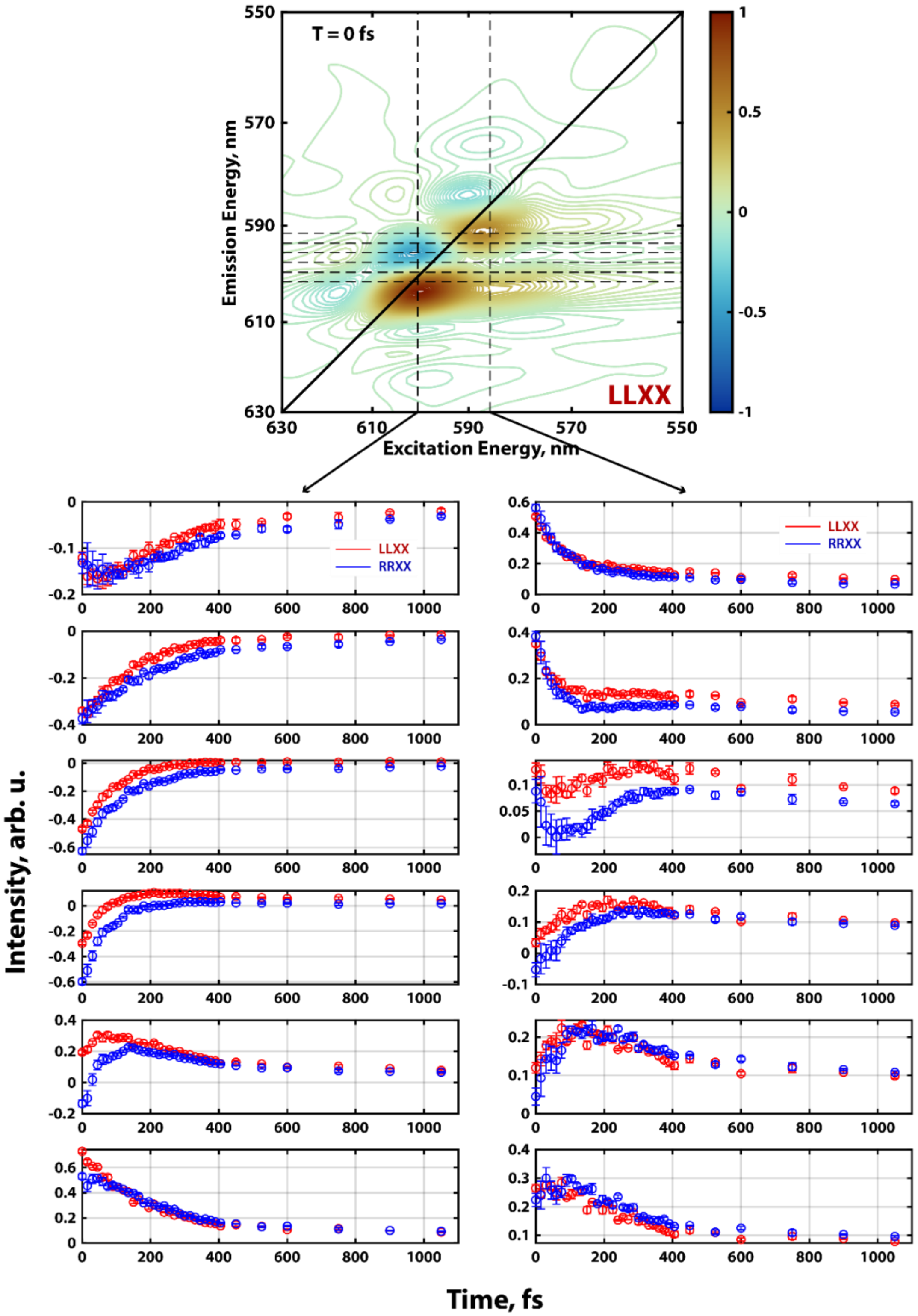

**Figure S11**: This figure shows that no growth of the below-diagonal positive cross-peak is seen in our data. The apparent "growth followed by decay" signal on the right side of the figure is due to mixing between positive and negative features. We plot the time traces on the left side for comparison to show that an apparent "growth and decay” can be seen as we move vertically from the inner tube ESA to the inner tube GSB (as we go down along the left vertical axis). The origin of this signal is due to cancellation between positive and negative features.

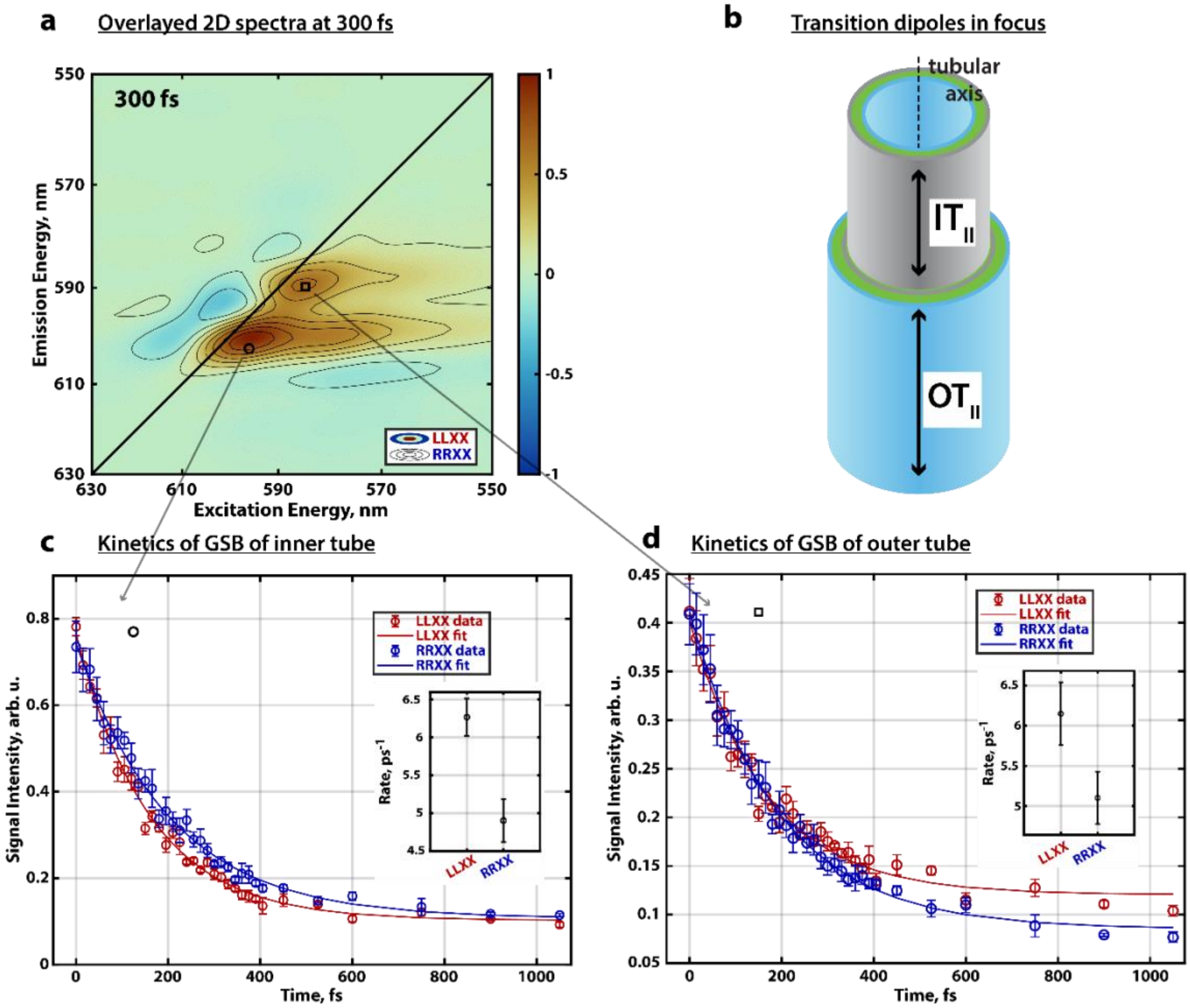

**Figure S12**: **(a)** Overlayed two-dimensional absorptive spectra at T = 300 fs of the ⟨LLXX⟩ (filled colored contours) and ⟨RRXX⟩ (black contour lines). **(b)** Schematic of the J-aggregate with the relevant transition dipoles for this figure. Time traces of different pulse sequences at **(c)** {$\lambda_{ex}$ = 595.5 nm, $\lambda_{em}$ = 602.5 nm}, (d) {$\lambda_{ex}$ = 583.9 nm, $\lambda_{em}$ = 589.6 nm}, Error bars on the time traces indicate standard error in the mean obtained from repeating each pulse sequence four times. The inset plots in parts **c** and **d** show time constants with standard errors in the mean obtained by fitting each repeated experiment separately.

## Statistical analysis – computing mean and standard error of decay rate constant: Figure 3

For a given pixel in the 2D spectrum, there are four replicate time traces available for each of the two pulse sequences, $\langle LLXX\rangle$ and $\langle RRXX\rangle$. We model the signal $s$ as an exponential decay with amplitude $A$ plus an intercept $c$.

$$s(t) = Ae^{-kt} + c \qquad \text{Equation S4}$$

For each pulse sequence, the four replicates are fit together in a joint (global) nonlinear regression to obtain a mean rate constant $k$ and its standard error. The obtained means and standard errors are plotted in the inset of **Figures 3** and **S12**. For example, for the $\langle LLXX\rangle$ sequence, replicate $i$ ($i = 1,2,3,4$) has a signal $S_{L,i}(t)$ measured at time points $t_1, t_2, \dots, t_{nT}$, modeled as

$$S_{L,i}(t) = A_{L,i}e^{-k_L t} + c_{L,i}$$

Equation S5

while swapping the suffix $L$ for $R$ for the $\langle RRXX\rangle$ sequence in **Equation S5**. For a given pulse sequence, the intercept and amplitude parameters are allowed to vary freely from replicate to replicate, absorbing shot-to-shot fluctuations and baseline differences, while the rate constant is written without a replicate index and is treated as a single property shared by all four replicates for that pulse sequence. The rationale behind this global fitting strategy over separately computing rates for the individual replicates *i* has been demonstrated before.(*12*, *13*) Briefly, when fitting a signal to an exponential decay, amplitude and rate constant are correlated parameters – a small increase in one can be partly offset by a change in another without much loss of fit quality. Therefore, an individual replicate's rate constant can be poorly constrained and be subject to the replicate's amplitude parameter, which is only subject to experimental noise.

The cost function to minimize is the sum of squared error (SSE), written for the $\langle LLXX\rangle$ experiment as

$$SSE_L(\theta_L) = \sum_{i=1}^{4}\sum_{j=1}^{nT}\left[s_{L,i} - A_{L,i}e^{k_L t_j} - c_{L,i}\right]^2 \qquad \text{Equation S6}$$

with $\theta_L = \{A_{L,1}, \dots, A_{L,4}, c_{L,1}, \dots, c_{L,4}, k_L\}$ is the parameter set. We use the Levenberg-Marquardt nonlinear least squares algorithm to minimize **Equation S6**, using the in-built MATLAB function "*lsqnonlin*". For a nonlinear least-squares fit, the parameter vector $\theta_L$ is approximately normally distributed about its true value, with a covariance matrix

$$Cov(\theta_L) = v_L^2(J_L^T J_L)^{-1} \qquad \text{Equation S7}$$

where $J_L$ is the Jacobian for the $\langle LLXX\rangle$ sequence's residuals with respect to $\theta_L$ evaluated at the best-fit solution, and $v_L^2$ is the residual variance estimated from the $\langle LLXX\rangle$ dataset.

$$v_L^2 = {SSE_L(\theta_L)}/{dof_L},\ dof_L = 4 \times nT - 9 \qquad \text{Equation S8}$$

This is the standard result for nonlinear least-squares estimation under approximately Gaussian errors.(*14*) The standard error in the mean for $k_L$ is the square root of the diagonal entry of $Cov(\theta_L)$. The identical set of computations is conducted on the $\langle RRXX\rangle$ dataset to obtain $k_R$ and its standard error in the mean.

**Statistical analysis – computing p-value for kinetics of $\langle LLXX\rangle$ and $\langle RRXX\rangle$: Figure 3**

This analysis answers whether two samples are drawn from the same parent population or from different populations. In our case, the specific question is whether time traces obtained from the $\langle LLXX\rangle$ and $\langle RRXX\rangle$ experiments contain the same information (drawn from the same population) or are they statistically different. The assumed model for our time traces is given by **Equation S4**, and the only parameter of interest is the rate constant $k$. Therefore, we design to test our hypothesis of whether $k_L = k_R$ (null hypothesis) while carefully handling other parameters like $A$ and $c$.

We employ a nested-model F-test.(*15*) The test for the null hypothesis conducts a fit for both pulse sequences to obtain a single rate constant and then conducts another fit where the rates are allowed to differ. The F-statistic is a quantification of whether allowing the rates to differ is statistically "significantly better" than forcing a single rate constant on both pulse sequence datasets, and consequently a p-value is obtained. Like calculating the standard error in the rate constants above, we allow the amplitude and intercept parameters to be separate between replicates but force one rate constant for a given pulse sequence for $k_L \neq k_R$(alternate hypothesis) or one rate constant for both pulse sequences for $k_L = k_R$(null hypothesis).

There are 34 waiting time points in each replicate dataset. There are 8 replicates in total (4 for each pulse sequence), implying 8 amplitude and 8 intercept parameters, and one rates for the null hypothesis and two rates for the alternate. Therefore, the total number of degrees of freedom for the null hypothesis is $8 \times 34 - (8 + 8 + 1) = 255$ and $254$ for the alternate hypothesis. Below, we show the analysis for the ESA pixel for inner tube ($\lambda_{ex}$=602.7 nm, $\lambda_{em}$=593.0 nm) shown in **Figure 3c**.

Computed sum of squared errors for the two hypotheses: $SSE_{null} = 0.09643$ and $SSE_{alt} = 0.08904$. Therefore, $F = {}^{(SSE_{null} - SSE_{alt})/(dof_{null} - dof_{alt})}\Big/_{SSE_{alt}/dof_{alt}} = {}^{0.09643 - 0.08904/1}\Big/_{0.08904/254} = 21.07.$

The p-value is given by the area under the $F(1{,}254)$ probability curve to the right of $F = 21.07$, which is $6.98 \times 10^{-6}$.

To compare the ESA spectral regions for the inner and outer nanotubes of the sample, we plot the p-values for every pixel in those spectral regions, as shown below.

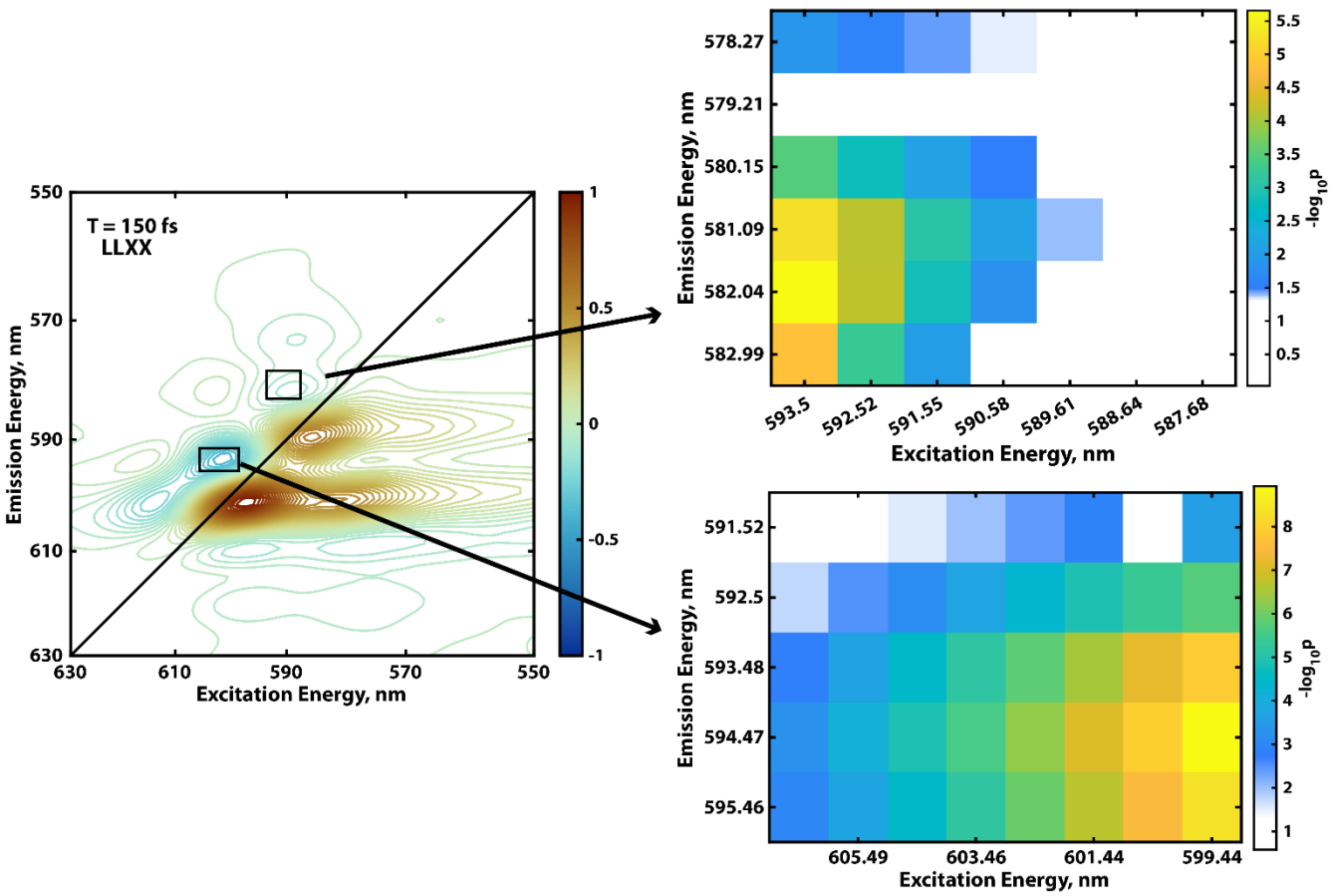

**Figure S13**: **Left:** 2D frame of the $\langle LLXX \rangle$ experiment at a waiting time of 150 fs is plotted with solid black squares inside the plot denoting the ESA spectral regions for the inner and outer tube. **Right:** Spectral regions of the inner and outer tube ESA features showing p-values obtained by comparing kinetics of each pixel inside those regions. Negative logarithm of the p-values is plotted. White pixels correspond to $p > 0.05$.

**Computing signal peak and signal rise rate constants: Figure 4**

The quantity of physical interest in these time traces is the magnitude of the transient – how large a signal the sample develops at the maximum of the rise. Our kinetic model is

$$s(t) = Ag(t) + c \qquad \text{Equation S9}$$

Where $g(t) = \left(1 - e^{-k_r t}\right)e^{-k_d t}$. To maximize $g(t)$, we set $g'(t) = 0$.

$$\frac{dg}{dt} = -k_d e^{-k_d t} + (k_r + k_d)e^{-(k_r+k_d)t} = 0 \Rightarrow e^{k_r t} = 1 + \frac{k_r}{k_d} \qquad \text{Equation S10}$$

$g'(t) = 0$ at $t = t_{peak}$, implying $t_{peak} = \frac{\ln(1+k_r/k_d)}{k_r}$. Plugging in $t_{peak}$ into $g(t)$, we get

$$g_{max} = \frac{1}{1+\rho}\left(\frac{\rho}{1+\rho}\right)^{\rho}, \quad \rho = \frac{k_d}{k_r} \qquad \text{Equation S11}$$

We want the maximum signal amplitude $A_{peak} = Ag_{max} + c$. Substituting this equation into Equation S9, we get (*16*).

$$s(t) = A(g(t) - g_{max}) + A_{peak} \quad \text{Equation S12}$$

**Explanation behind observed difference in cross-peak rise amplitude: Figure 4**

Despite the absence of chiral dynamics of the inner tube transverse exciton, we observe a rise of the cross-peak intensity with a higher amplitude in the $\langle LLXX \rangle$ and $\langle RRXX \rangle$ experiments compared to the $\langle XXXX \rangle$ experiment (see **Figure 4D**). We denote this maximum signal amplitude as $A_{peak}$, and derive an analytical expression for $A_{peak}$ starting from the rise-decay kinetic model in **Equation S12** (*41*). We present the data and the fitted values of $A_{peak}$ in **Figure 4D** and observe that $\frac{A_{peak}(LLXX)}{A_{peak}(XXXX)} \approx 1.45$, and $\frac{A_{peak}(RRXX)}{A_{peak}(XXXX)} \approx 1.57$. To understand this observation, it is instructive to decompose a pulse sequence with both linear and circularly polarized pulses into linear combinations of sequences constituting only linearly polarized pulses, as shown below (*42*).

$$\langle LLXX \rangle = \frac{1}{2}[\langle XXXX \rangle + \langle YYXX \rangle \pm i(\langle XYXX \rangle - \langle YXXX \rangle)]$$

$$\langle RRXX \rangle = \frac{1}{2}[\langle XXXX \rangle + \langle YYXX \rangle \mp i(\langle XYXX \rangle - \langle YXXX \rangle)] \quad \text{Equation S13}$$

In **Equation S13**, the sign before the imaginary term is '+' for rephasing pathways and '−' for non-rephasing pathways for $\langle LLXX \rangle$ and vice-versa for $\langle RRXX \rangle$, and is discussed in **Section 4**. Note that the third and fourth terms of **Equation S13** have their signs flipped between $\langle LLXX \rangle$ and $\langle RRXX \rangle$ experiments and are therefore responsible for any differences arising due to chirality (*33*, *34*). Since we do not observe chiral dynamics for the transverse exciton, we invoke the electric dipole approximation and ignore the third and fourth terms of **Equation S13**. The energy transfer response function between orthogonal dipoles for the polarization sequence $\langle YYXX \rangle$ has an orientationally averaged pre-factor of $2/15$, while that for $\langle XXXX \rangle$ is $1/15$ (*43*). The pre-factors add up to $\frac{1}{2} \times \left(\frac{1}{15} + \frac{2}{15}\right) = \frac{1}{10}$ for both $\langle LLXX \rangle$ and $\langle RRXX \rangle$, thereby yielding

$$\frac{A_{peak}(LLXX)}{A_{peak}(XXXX)} = \frac{A_{peak}(RRXX)}{A_{peak}(XXXX)} = \frac{1/10}{1/15} = 1.5 \quad \text{Equation S14}$$

Our observed peak ratios of 1.45 and 1.57 are in good agreement with **Equation S14**, thus explaining the observed difference in maximum signal amplitude between the pulse sequences.

## 6. Additional Spectroscopy Data

To ensure the robustness of our observations, we conducted the experiments in the following order in lab time: Nile Blue $\langle LLXX \rangle$ → C8S3 nanotubes $\langle LLXX \rangle$ → C8S3 nanotubes $\langle XXXX \rangle$ → Nile Blue $\langle XXXX \rangle$ → C8S3 nanotubes $\langle RRXX \rangle$ → Nile Blue $\langle RRXX \rangle$. This sequence of experiments ensures that any differences in dynamics that may be attributed to any manner of laser drift (amplitude or phase) as a function of lab time between the left- and right-handed pumps will also show up in the Nile Blue control. It also ensures that the control experiments are conducted with the same exact optical alignment as the experiments involving chiral nanotubes.

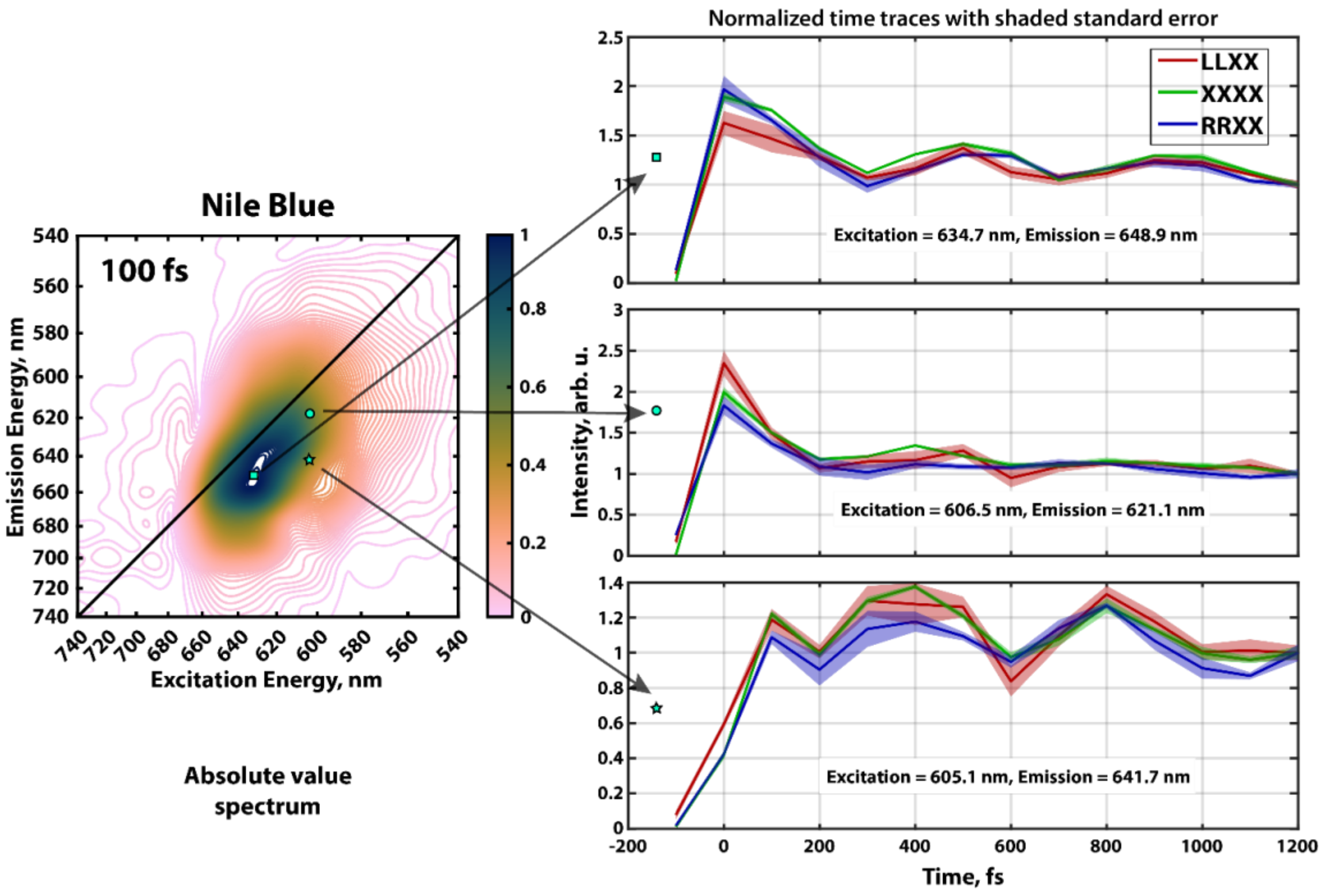

**Figure S14**: Control experiment on Nile Blue demonstrating no change in nonlinear dynamics as a function of the pump pulse polarization. Time traces on the right side of the figure are normalized to the last point. Shaded region around the time traces denotes one standard error in the mean.

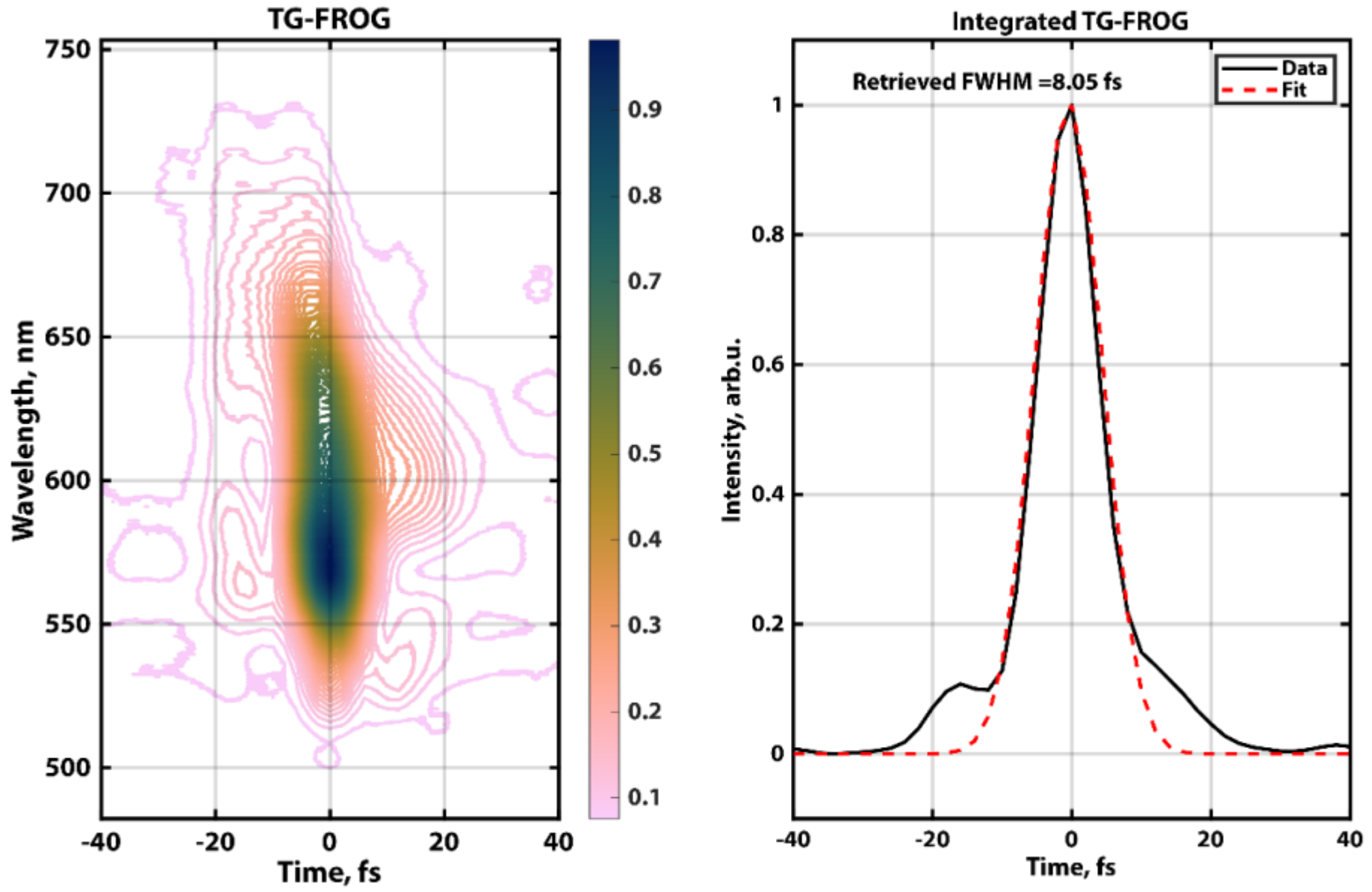

**Figure S15**: Transient Grating Frequency Resolved Optical Gating (TG-FROG) autocorrelation measurement as a function of pump-probe time delay (abscissa) on glass. Left figure shows the autocorrelation intensity spectrogram. Right figure shows a Gaussian fit to the signal integrated over the wavelength axis.

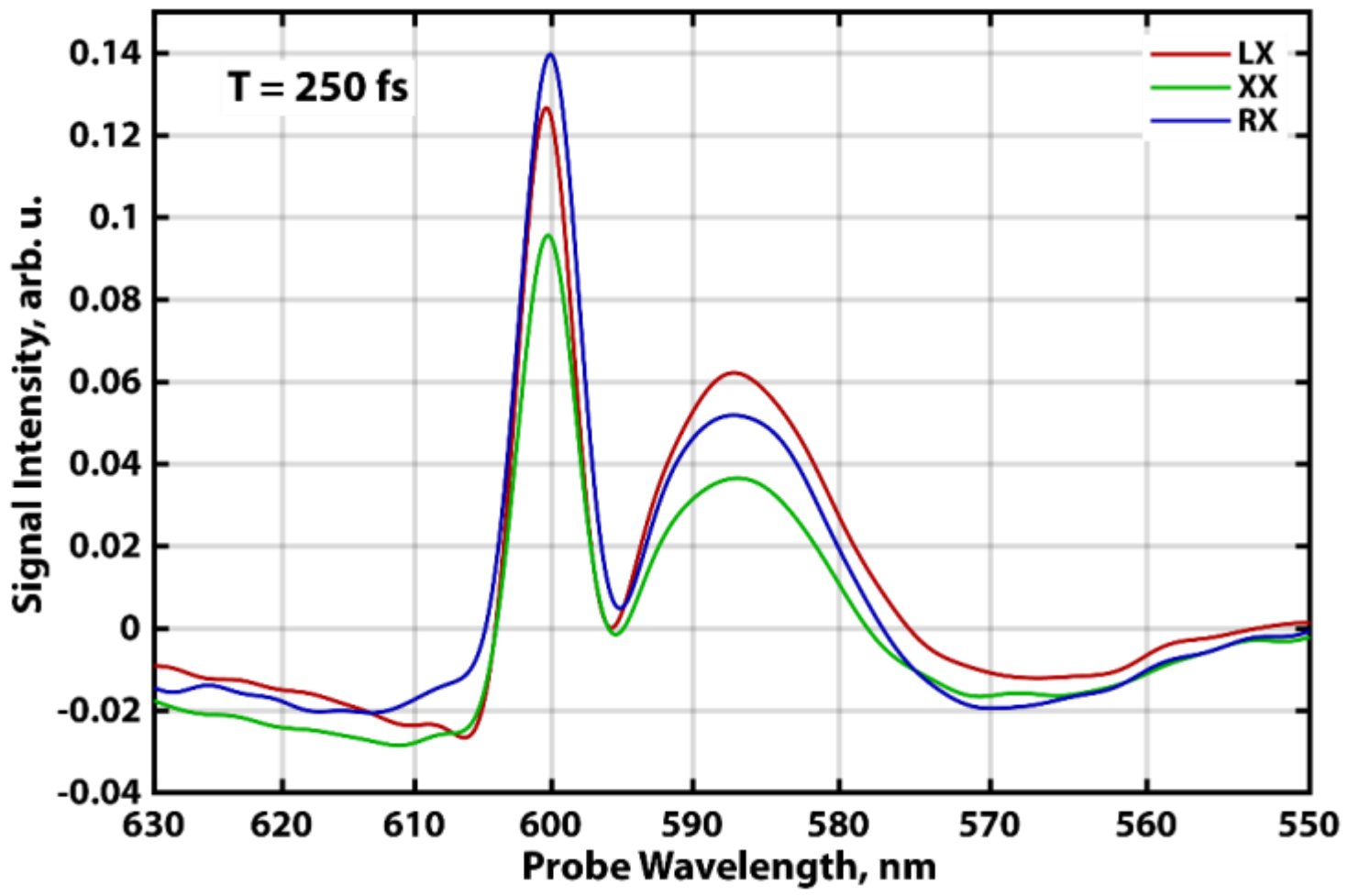

**Figure S16**: Transient absorption spectra of the C8S3 chiral aggregate for pulse sequences LX, XX, and RX at a waiting time of 250 fs used for phasing the 2D spectra.

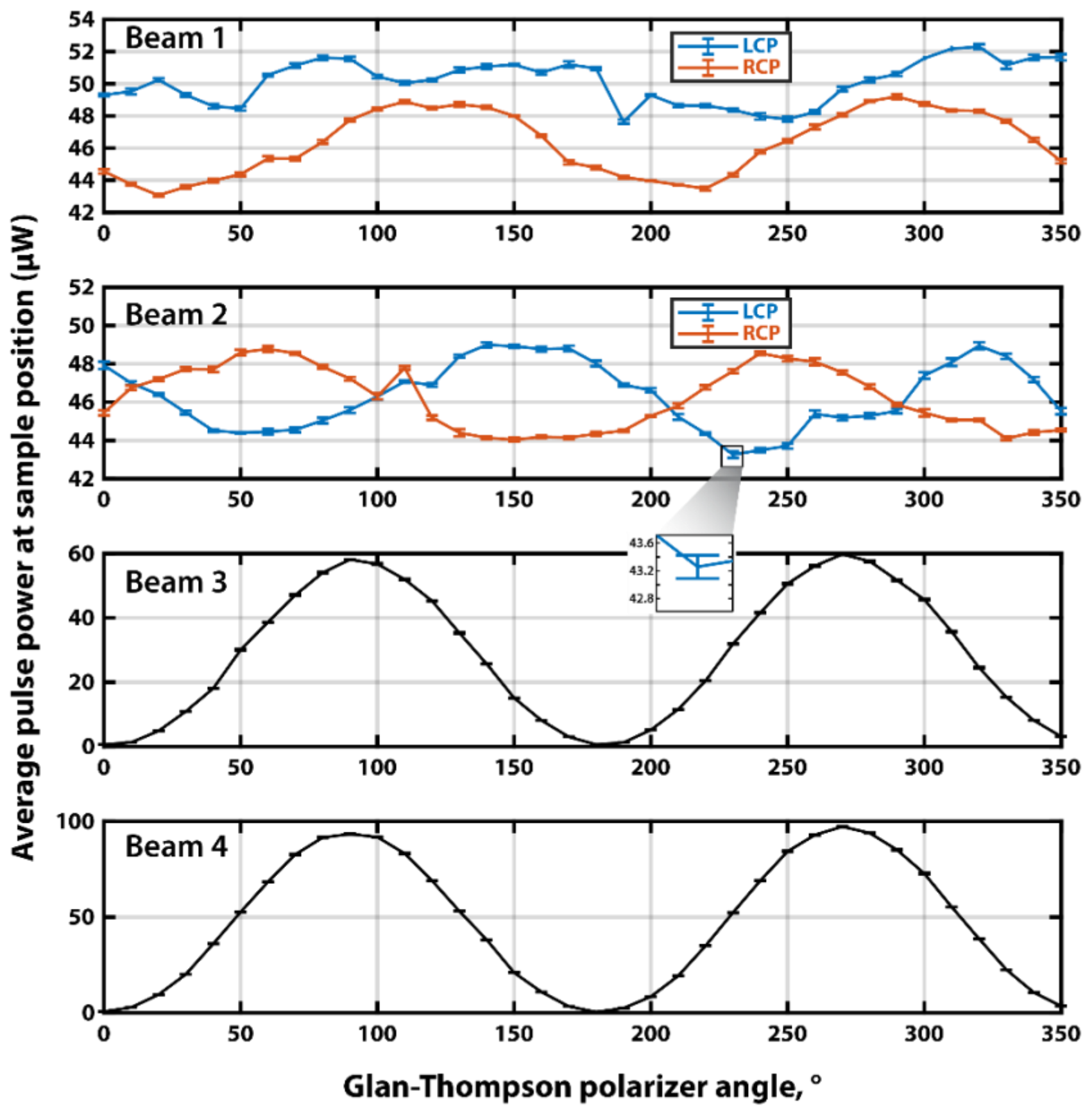

**Figure S17**: Power of four pulses recorded at the sample position right after a Glan-Thompson polarizer.